\documentclass[notitlepage, reprint, amsmath, amssymb, aps, onecolumn]{revtex4-1}

\usepackage{amssymb}
\usepackage{hyperref}
\usepackage{bm}
\usepackage{soul}
\usepackage{ulem}
\usepackage{dsfont}
\usepackage{amsthm}
\usepackage{amsmath}
\usepackage{verbatim} 
\usepackage{qcircuit}
\usepackage{circuitikz}
\usepackage{tabularx}
\usepackage{color,soul}
\usepackage{esint}
\usepackage{physics}
\usepackage{braket}
\usepackage{bbold}
\usepackage{amssymb}
\usepackage{float}
\usepackage{subfigure}

\usepackage{todonotes}

\begin{document}

\author{A.~S. Kardashin}
\email{andrey.kardashin@skoltech.ru}
\affiliation{Skolkovo Institute of Science and Technology, Moscow 121205, Russia}
\author{A.~V. Vlasova, A.~A. Pervishko, D.~Yudin}
\affiliation{Skolkovo Institute of Science and Technology, Moscow 121205, Russia}
\author{J.~D. Biamonte}
\homepage{http://quantum.skoltech.ru}
\affiliation{Skolkovo Institute of Science and Technology, Moscow 121205, Russia}

\title{Quantum machine learning channel discrimination}

\begin{abstract}
    In the problem of quantum channel discrimination, one distinguishes between a given number of quantum channels, which is done by sending an input state through a channel and measuring the output state. 
    This work studies applications of variational quantum circuits and machine learning techniques for discriminating such channels.
    In particular, we explore (i) the practical implementation of embedding this task into the framework of variational quantum computing, (ii) training a quantum classifier based on variational quantum circuits, and (iii) applying the quantum kernel estimation technique.
    For testing these three channel discrimination approaches, we considered a pair of entanglement-breaking channels and the depolarizing channel with two different depolarization factors.
    For the approach (i), we address solving the quantum channel discrimination problem using widely discussed parallel and sequential strategies.
    We show the advantage of the latter in terms of better convergence with less quantum resources.
    Quantum channel discrimination with a variational quantum classifier (ii) allows one to operate even with random and mixed input states and simple variational circuits.
    The kernel-based classification approach (iii) is also found effective as it allows one to discriminate depolarizing channels associated not with just fixed values of the depolarization factor, but with ranges of it.
    Additionally, we discovered that a simple modification of one of the commonly used kernels significantly increases the efficiency of this approach.
    Finally, our numerical findings reveal that the performance of variational methods of channel discrimination depends on the trace of the product of the output states.  These findings demonstrate that quantum machine learning can be used to discriminate channels, such as those representing physical noise processes.   
\end{abstract}

\maketitle
    
\section{Motivation}
    
    A quantum channel is a linear completely-positive trace-preserving map which transforms quantum states into quantum states.
    The problem of quantum channel discrimination, i.e. distinguishing between a given finite set of channels, is ubiquitous in quantum information and quantum communication~\cite{kitaev1997quantum, wang2006unambiguous, duan2009perfect,Holevo_2012}. 
    Solving this task forms the core of various applications, including but not limited to photonic sensing~\cite{pirandola2018advances}, target quantum  detection via quantum illumination~\cite{lloyd2008enhanced, barzanjeh2015microwave}, and quantum reading~\cite{ortolano2021experimental}. 
    Within a general approach for implementing quantum channel discrimination one sends an input state as specified by its density operator $\rho^\mathrm{in}$ through a channel $\Phi_y$ randomly selected from a collection $\{\Phi_j\}_{j=1}^N$.
    The output state, $\rho^\mathrm{out} = \Phi_y[\rho^\mathrm{in}]$, may be equivalently expressed in terms of positive operator-valued measures (POVM)~\cite{Brandt_1999}. 
    The POVM is constituted by a set of non-negative Hermitian operators $\Pi = \{\Pi_j\}_{j=1}^N$  that add up to the identity. 
    The probability for the quantum system to be in a particular state $\Phi_y[\rho]$ is determined by the expectation value of the POVM operator corresponding to this state $\Tr(\Pi_y\Phi_y[\rho^\mathrm{in}])$~\cite{James_2001,Palmieri_2020}. 
    Wheres quantum channel discrimination should return the index $y$ which labels the channel having been applied.
    
    Depending on the available computational resources and the properties of a given channel $\Phi_y$, one might adopt various strategies for discrimination~\cite{harrow2010adaptive, wilde2020coherent, farooq2018quantum, bavaresco2021strict}. 
    In particular, it is expected that quantum channels can be discriminated more efficiently if it is allowed to apply a given channel several times. 
    For different channels and discrimination strategies, the efficiency of discrimination as well as the associated quantitative measures have previously been considered~\cite{cooney2016strong, katariya2021geometric, Pirandola_2019, pereira2021bounds, zhuang2020ultimate, wilde2020amortized, katariya2021evaluating, salek2020adaptive}. 
    Of particular interest and technological relevance \cite{Harney_2021} are the parallel and sequential strategies~\cite{Harney_2021}.
    Given the opportunity to apply a channel a fixed number of times, the parallel strategy implies that a channel is applied on distinct quantum subsystems (e.g. qubits) simultaneously, wheres in the sequential strategy a channel acts on the same subsystem step by step.
    
    The problem of quantum channel discrimination can be seen as an optimization task in the space of parameters specified by the input state $\rho^\mathrm{in}$ and the POVM $\Pi$. 
    This allows one to put this task into the framework of variational quantum computing~\cite{mcclean2016theory, li2017hybrid, santagati2018witnessing, peruzzo2014variational, biamonte2021universal, cerezo2021variational, larose2019variational, lubasch2020variational} and quantum machine learning~\cite{biamonte2017quantum, huggins2019towards, schuld2015introduction, schuld2021machine, schuld2019quantum}.
    In the variational scheme, a quantum processor is used to prepare a family of probe states with a polynomial number of parameters, while minimizing a given loss function within this family of states is achieved by a means of classical optimization algorithms. 
    In this case, one minimizes the use of quantum resources by delegating as much computation as possible to the classical device.
    Machine learning as implemented with variational quantum circuits~\cite{mitarai2018quantum, schuld2020circuit, chen2020variational} allows one to solve classification tasks for quantum-embedded classical data~\cite{lloyd2020quantum}, learn phases of quantum matter~\cite{uvarov2020machine,Uvarov_2020,Kardashin_2020,Kardashin_2021,lazzarin2022multi}, and discriminate quantum states~\cite{patterson2021quantum, chen2021universal}.
    Two latter tasks are peculiar in a sense that they are solved for quantum data by quantum means, which makes them a part of the rapidly advancing quantum-quantum machine learning~\cite{mengoni2019kernel, yu2019reconstruction}, and the problem of channel discrimination can be classified likewise.
   
    In this Paper, we highlight the use of variational quantum circuits and machine learning techniques for quantum channel discrimination.
    We start our analysis with formulating this task as an optimization problem, followed by a direct application of variational quantum scheme. 
    Then we discuss how one can use variational circuits for binary classification of quantum channels.
    After that, we demonstrate that such a binary classifier can be also based on a kernel, a specific real-valued function of output states of channels, $\mathcal{K}(\Phi_i[\rho],\Phi_j[\rho])$ for an input state $\rho$.
    We test these three methods of channel discrimination by distinguishing between a pair of entanglement-breaking channels~\cite{horodecki2003entanglement, ruskai2003qubit}, and the qubit depolarizing channel~\cite{shaham2015entanglement, shaham2012realizing}.

\section{Quantum channel discrimination}
\label{sec:quantum_channel_discrimination}
    
    The problem of binary quantum channel discrimination can equivalently be viewed as the game of two parties, Alice and Bob.
    This game is consisted of the active stage and the training stage.
    At the beginning of the active stage, Alice prepares the state $\rho^\mathrm{in}$ and sends it to Bob.
    Then, Bob, in secret, randomly and with equal probabilities chooses a channel $\Phi_y \in \{\Phi_0, \Phi_1\}$ and applies it to the Alice's state, so that the output state is $\rho^\mathrm{out} = \Phi_y[\rho^\mathrm{in}]$.
    Having received this state from Bob, Alice measures the output state with the POVM elements $\Pi = \{ \Pi_0, \Pi_1 = \mathbb{1} - \Pi_0 \}$, and label the outcome `0' for the channel $\Phi_0$ and `1' otherwise.
    The goal of Alice therefore is to find the input state $\rho^\mathrm{in}$ and the POVM elements $\Pi = \{ \Pi_0, \Pi_1 \}$ such that they maximize the probability $\mathrm{p}_{00}$ of getting the measurement outcome `0' if Bob applied the channel $\Phi_0$ and the probability $\mathrm{p}_{11}$ of getting the outcome `1' if the applied channel was $\Phi_1$.
    These probabilities are 
    \begin{align*}
        \mathrm{p}_{00} &= \frac{1}{2} \Tr(\Pi_0 \Phi_0[\rho^\mathrm{in}]), \\
        \mathrm{p}_{11} &= \frac{1}{2} \Tr(\Pi_1 \Phi_1[\rho^\mathrm{in}]),
    \end{align*}
    where the factors $1/2$ are essentially the prior probabilities of application of the channel $\Phi_0$ or $\Phi_1$ since Bob chooses them randomly and equiprobably.
    As $\mathrm{p}_{00} + \mathrm{p}_{11}$ forms the total probability of successful discrimination, then the task of Alice is to maximize it.
    This can formally described as the following optimization problem:
    \begin{equation}\label{eq:p_suc}
        \mathrm{p_{s}} \equiv \max\{\mathrm{p}_{00} + \mathrm{p}_{11}\} =  \frac{1}{2} \max_{\rho^\mathrm{in}, \Pi} \big\{ \Tr(\Pi_0 \Phi_0[\rho^\mathrm{in}]) + \Tr(\Pi_1 \Phi_1[\rho^\mathrm{in}]) \big\}.
    \end{equation}
    Alongside with the probability of erroneous discrimination $\mathrm{p_{e}} = \mathrm{p}_{01} + \mathrm{p}_{10} $, it obviously sums to unity, $\mathrm{p_{s}} + \mathrm{p_{e}} = 1$.
    
    To be able to solve the problem \eqref{eq:p_suc} for Alice, there is the training stage in the game.
    During this stage, we assume that for each pair $(\rho^\mathrm{in}, \Pi)$ picked by Alice, Bob provides as many copies of the state $\rho^\mathrm{out} = \Phi_y[\rho^\mathrm{in}]$ as needed, without changing the label $y\in\{0, 1\}$.
    Moreover, for each output state, Alice is informed about the channel label $y$.
    This is equivalent to the assumption that Alice can measure the output state $\rho^\mathrm{out}$ arbitrarily many times, i.e. it is possible to compute the probabilities in \eqref{eq:p_suc} exactly.
    This is opposed to the situation at the game's active stage, when only one measurement is allowed, and the chanel label $y$ is kept in secret.
    Clearly, the described game scheme is analogous to binary classification task as realized in the context of supervised machine learning.
    
    In this particular setting, the probability of successful quantum channel discrimination that can be achieved is upper bounded by \cite{benenti2010computing}
    \begin{equation} \label{eq:p_suc_diamond}
        \mathrm{p}_\diamond = \frac{1}{2} - \frac{1}{4} ||\Phi_0 - \Phi_1||_\diamond,
    \end{equation}
    where
    \begin{equation*}
        ||\Phi||_\diamond = \max_{\rho} \big|\big| (\Phi \otimes \mathbb{1})[\rho]\big|\big|_1
    \end{equation*}
    is the so-called diamond-norm with $\Phi$ being a channel that maps density operators on a Hilbert space $\mathcal{H}$, $\mathbb{1}$ the identity map on $\mathcal{H}_E$, $\rho$ a density operator on $\mathcal{H} \otimes \mathcal{H}_E$, and $||A||_1 = \Tr \sqrt{A^\dagger A}$.

\subsection{Discrimination strategies}
    
    So far we have described a channel guessing game when a channel $\Phi_y$ is applied only once.
    If however Alice is allowed to pass a chosen probe state $\rho^\mathrm{in}$ through the Bob's channel $\Phi_y$ a finite and fixed number of times $p$, provided that for each time the channel label $y$ remains the same, then Alice can adjust the discrimination strategy by asking Bob to apply the channel $\Phi_y$ in a specific way.
    One of such ways is applying the channel $p$ times in parallel, so that the channel acts simultaneously on the separate subsystems of a composite input state $\rho^{\mathrm{in}}$.
    Another approach is to apply the channel sequentially such that it acts on a single subsystem of a potentially composite state $\rho^{\mathrm{in}}$.
    These two discrimination methods are widely discussed in literature and are called the parallel and sequential strategies, respectively.
    In what follows, we formally describe these strategies assuming that the channels $\Phi_y$ are qubit-to-qubit mappings.
    
    \subsubsection{Parallel strategy}
    
\begin{figure}[H]
        \centering
        \includegraphics{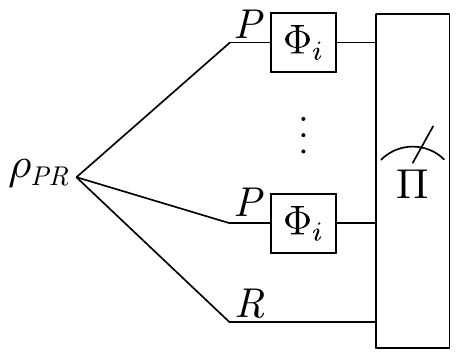}
        \caption{
        A schematic of the parallel strategy.
        Provided with the opportunity of $p$ applications of the Bob's channel $\Phi_y$, Alice prepares a composite state $\rho_{PR}^\mathrm{in}$ of $p$ qubits of the register $P$ and $r$ qubits of the register $R$.
        The qubits of $P$ are sent through the channels $\Phi_y$, while the qubits of $R$ remain untouched.
        In the scheme, the lines coming from $\rho_{PR}^\mathrm{in}$ indicate the subsystems of the corresponding registers.
        At the end, all $p+r$ qubits are measured with the POVM $\Pi$.
        }
        \label{fig:parallel_schem}
    \end{figure}

    Consider the previously discussed quantum channel discrimination game and suppose that now Alice can ask Bob to apply the channel $\Phi_y$ a fixed number of times $p$.
    In general words then, the parallel channel discrimination strategy implies that the input state of Alice $\rho^\mathrm{in}$ is at least $p$-qubit, and Bob acts by the channels $\Phi_y$ on each of the $p$ qubits separately and simultaneously.
    Alice is also allowed to have the input state of more than $p$ qubits, i.e. to add $r \geqslant 0$ auxiliary qubits to it, as it might potentially help in solving quantum channel discrimination task in case of an entangled input state~\cite{farooq2018quantum, puzzuoli2017ancilla, matthews2010entanglement}.
    The resultant $(p+r)$-qubit state is then measured with a POVM $\Pi$.
    This discrimination strategy is schematically shown in Fig.~\ref{fig:parallel_schem}
    
    Let us describe the parallel strategy more formally.
    Suppose that the channel $\Phi_y$ can be applied $p$ times.
    Then, first, Alice prepares a $(p+r)$-qubit state $\rho_{PR}^\mathrm{in}$, with $P$ and $R$ specifying the registers of $p$ and $r$ qubits, respectively.
    The qubits of $P$ are then sent through the channels of Bob in parallel, while the register $R$ remains unaffected.
    Formally, the output state is 
    \begin{equation} \label{eq:out_state_par}
        \rho^\mathrm{out} = (\Phi_y^{\otimes p} \otimes \mathbb{1}^{\otimes r})[\rho_{PR}^\mathrm{in}],
    \end{equation}
    where $\Phi_y^{\otimes p}$ acts only on the subsystem $P$ and $\mathbb{1}^{\otimes r}$ is the identity on the subsystem $R$.
    Alice then measures the output state $\rho^\mathrm{out}$ with $\Pi= \{\Pi_0, \Pi_1 \}$. Therefore, in the parallel strategy, according to Eq.~\eqref{eq:p_suc} one has to maximize
    \begin{equation} \label{eq:p_suc_par}
        \mathrm{p_s^{par}} = \frac{1}{2} \max_{\rho_{PR}^\mathrm{in}, \Pi} \big\{ \Tr(\Pi_0 (\Phi_0^{\otimes p} \otimes \mathbb{1}^{\otimes r})[\rho_{PR}^\mathrm{in}]) + \Tr(\Pi_1 (\Phi_1^{\otimes p} \otimes \mathbb{1}^{\otimes r})[\rho_{PR}^\mathrm{in}]) \big\},
    \end{equation}
    where the optimization over $r$, specifying the number of qubits in the register $R$, is implicitly included.
    As was mentioned earlier, introducing this auxiliary register allows to have the entanglement between the qubits of the registers $P$ and $R$, which may lead to more efficient channel discrimination.
    For the described strategy, the probability of successful channel discrimination is yielded by
    \begin{equation}  \label{eq:p_suc_diamond_par}
        \mathrm{p}_\diamond^\mathrm{par}(p) = \frac{1}{2} - \frac{1}{4} ||\Phi_0^{\otimes p} - \Phi_1^{\otimes p}||_\diamond
    \end{equation}
    for $p$ parallel applications of the Bob's channel.
    
    \subsubsection{Sequential strategy}
    
\begin{figure}[h]
        \centering
        \includegraphics{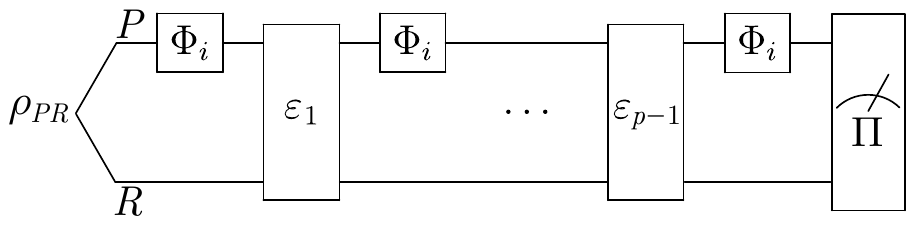}
        \caption{
        A schematic of the sequential strategy.
        Provided with the opportunity of $p$ applications of the Bob's channel $\Phi_y$, Alice prepares the composite state $\rho_{PR}^\mathrm{in}$ of one qubit of the register $P$ and $r$ qubits of the register $R$.
        The qubit of $P$ is sent through the channels $\Phi_y$, while the qubits of $R$ remain untouched.
        After the $j$th application of the channel $\Phi_y$, Alice modifies the whole state by applying the channel $\mathcal{E}_j$.
        In the scheme, the lines coming from $\rho_{PR}^\mathrm{in}$ indicate the subsystems of the corresponding registers.
        At the end, all $1+r$ qubits are measured with the POVM $\Pi$.
        }
        \label{fig:sequential_schem}
    \end{figure}

    Now suppose that Alice is again allowed to apply the channel $\Phi_y$ a fixed number of times $p$.
    This time, in the sequential discrimination strategy, the Alice's input state $\rho^\mathrm{in}$ can be single-qubit, and this qubit can be passed through a channel $p$ times in a row.
    But after each application of the channel $\Phi_y$, Bob sends the corresponding output state back to Alice, who is allowed to modify it before sending back to Bob again.
    After the $p$th channel application, the resultant state is measured with the POVM $\Pi$.
    Like in the parallel strategy, Alice can add $r$ auxiliary qubits to have a $(1+r)$-qubit entangled input state.
    The sequential channel discrimination strategy is schematically shown in Fig.~\ref{fig:sequential_schem}.
    
    Formally, this strategy can be described as follows.
    First, Alice prepares the input state $\rho_{PR}^\mathrm{in}$ which consists of $(1+r)$ qubits: the subsystem $P$ of one qubit is acted by the channel $\Phi_y$, whereas the subsystem $R$ of $r$ qubits remains unaffected.
    Suppose Alice is allowed to apply the channel $p$ times.
    Alice then sends the input state $\rho_{PR}^\mathrm{in}$ to Bob and receives back the state $\tilde{\rho}_{PR} = (\Phi_y\otimes \mathbb{1}^{\otimes r})[\rho_{PR}^\mathrm{in}]$.
    After that, Alice applies a quantum channel $\varepsilon_1$ on the state $\tilde{\rho}_{PR}$ which gives $\rho_1 = \varepsilon_1[\tilde{\rho}_{PR}]$. 
    This procedure is repeated $(p-1)$ times until Alice has the state $\rho_{p-1} = \varepsilon_{p-1}[\rho_{p-2}]$, and at the end Alice passes the subsystem $P$ through the channel $\Phi_y$ the last, $p$th time, and gets $\rho^\mathrm{out} = (\Phi_y\otimes \mathbb{1}^{\otimes r})[\rho_{p-1}]$. More formally, the whole process can be described as
    \begin{equation} \label{eq:out_state_seq}
        \rho^\mathrm{out}
        = \mathcal{C}\big(\Phi_y, \mathcal{E}\big) [\rho_{PR}^\mathrm{in}], 
    \end{equation}
    where $\mathcal{E} = \{ \varepsilon_j \}_{j=1}^{p-1}$ and
    \begin{equation*}
        \mathcal{C}\big(\Phi_y, \mathcal{E} \big) = (\Phi_y \otimes \mathbb{1}^{\otimes r}) \circ \varepsilon_{p-1} \circ \cdots \circ (\Phi_y\otimes \mathbb{1}^{\otimes r}) \circ \varepsilon_{2} \circ (\Phi_y \otimes \mathbb{1}^{\otimes r}) \circ \varepsilon_{1} \circ (\Phi_y \otimes \mathbb{1}^{\otimes r})
    \end{equation*}
    with the channel composition operation  $(B \circ A) [\rho] \equiv B\big[A [\rho]\big]$.
    The output state $\rho^\mathrm{out}$ is then measured using the POVM $\Pi = \{\Pi_0, \Pi_1\}$.
    In this strategy, the optimization problem \eqref{eq:p_suc} becomes
    \begin{equation} \label{eq:p_suc_seq}
        \mathrm{p_s^{seq}} = \frac{1}{2} \max_{\rho_{PR}^\mathrm{in}, \Pi, \mathcal{E}} \big\{\Tr(\Pi_0\,\mathcal{C}(\Phi_0, \mathcal{E})[\rho_{PR}^\mathrm{in}])+\Tr(\Pi_1\, \mathcal{C}(\Phi_1, \mathcal{E})[\rho_{PR}^\mathrm{in}])\big\},
    \end{equation}
    where in addition to the input state $\rho_{PR}^\mathrm{in}$ and measurement $\Pi$, Alice has to optimize over the channels $\mathcal{E}$ as well.
    
    As a rule, the sequential strategy, incarnating the idea of quantum comb \cite{pirandola2018advances, chiribella2008quantum}, provides better discrimination results compared to the parallel strategy \cite{salek2020adaptive, katariya2021evaluating}. Meanwhile, it is worth mentioning a more general discrimination strategy that is based on the so-called indefinite casual order of channel application \cite{bavaresco2021strict}. It was shown that for $\mathrm{p^{par}}$, $\mathrm{p^{seq}}$, and $\mathrm{p^{ico}}$ being the upper bounds for successful discrimination probabilities of the parallel, sequential, and indefinite casual order strategies, respectively, there exists a pair of target channels $\Phi_0$ and $\Phi_1$ satisfying
    \begin{equation*}
        \mathrm{p^{par}} < \mathrm{p^{seq}} < \mathrm{p^{ico}}.
    \end{equation*}
    Although the indefinite casual order strategy gives advantage over the parallel and sequential ones, we focus on the latter two in what follows.
    
\subsection{Variational circuit formulation}
    
    Let us embed the parallel and sequential discrimination strategies into the framework of variational quantum circuits. 
    That is, we replace all transformations of quantum states by parametrized unitary operators.
    Having done that, we accordingly reformulate the optimization problems for $\mathrm{p_s^{par}}$ and $\mathrm{p_s^{seq}}$ defined in \eqref{eq:p_suc_par} and \eqref{eq:p_suc_seq}, respectively.
    For the case of a single channel application ($p=1$), a similar approach of embedding was applied for discriminating various qubit-to-qubit channels~\cite{qiu2019solving}.
    
    \subsubsection{Parallel strategy}
    
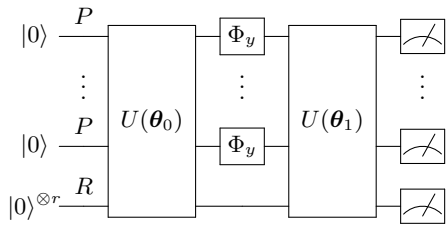
\begin{figure}[h]
        \centering
        \mbox{
            $$
            \Qcircuit @C=1.0em @R=1.0em {
                & \ket{0}             & & \ustick{P}\qw & \multigate{4}{U(\boldsymbol{\theta}_0)} & \gate{\Phi_y} & \multigate{4}{U(\boldsymbol{\theta}_1)} & \meter \\
                &                     & & \vdots        &                                         & \vdots        &                                         & \vdots \\
                &                     & &               &                                         &               &                                         &        \\
                & \ket{0}             & & \ustick{P}\qw & \ghost{U(\boldsymbol{\theta}_0)}        & \gate{\Phi_y} & \ghost{U(\boldsymbol{\theta}_1)}        & \meter \\
                & \ket{0}^{\otimes r} & & \ustick{R}\qw & \ghost{U(\boldsymbol{\theta}_0)}        & \qw           & \ghost{U(\boldsymbol{\theta}_1)}        & \meter \\
            }
            $$
        }
        \caption{
        A variational quantum circuit implementing the parallel channel discrimination strategy from Fig.~\ref{fig:parallel_schem}.
        The initial state $\rho_{PR}^\mathrm{in}$ is prepared from $\ket{0}^{\otimes p}_P \otimes \ket{0}^{\otimes r}_R$ via the unitary transformation $\mathcal{U}(\boldsymbol{\theta_0})$, while the unitary $\mathcal{U}(\boldsymbol{\theta_1})$ is used to rotate the measurement basis.
        If the number of allowed applications of the channel $\Phi_y$ is $p$ and the number of ancillary qubits is $r$, the technique requires $(p+r)$ qubits.
        Note that $r$ could be set to zero.}
        \label{fig:par_circ_var}
    \end{figure}

    The parallel channel discrimination strategy \eqref{eq:p_suc_par} embedded into the framework of variational circuits is depicted in Fig.~\ref{fig:par_circ_var}.
    In this circuit, the probe state $\rho_{PR}^\mathrm{in}$ is prepared as 
    \begin{equation*}
        \rho_{PR}^\mathrm{in} = \mathcal{U}(\boldsymbol{\theta}_0)[\rho_0(p, r)],
    \end{equation*}
    where $\rho_0(p, r) = \ketbra{0}{0}^{\otimes p}_P \otimes \ketbra{0}{0}^{\otimes r}_R$, $\mathcal{U}(\boldsymbol{\theta}_0)[\rho] = U(\boldsymbol{\theta}_0) \rho U^\dagger(\boldsymbol{\theta}_0)$, and $U(\boldsymbol{\theta}_0)$ is a unitary operator parametrized by a set of real numbers $\boldsymbol{\theta}_0$.
    The register $P$ of $p$ qubits for $\rho_{PR}^\mathrm{in}$ is acted then by the $p$ parallel applications of the channel $\Phi_y$, as it is done in~\eqref{eq:out_state_par}.
    After that, applied is the unitary $U(\boldsymbol{\theta}_1)$, which can be viewed as a rotation of the measurement basis.
    The output state then becomes
    \begin{equation} \label{eq:out_state_par_var}
        \rho^\mathrm{out}(\boldsymbol{\theta}, \Phi_y, p, r)
         = \mathcal{U}(\boldsymbol{\theta_1}) \circ \mathcal{F}(\Phi_y, p, r) \circ \mathcal{U}(\boldsymbol{\theta_0})[\rho_0(p, r)],
    \end{equation}
    where $\boldsymbol{\theta} = \boldsymbol{\theta}_0 \cup \boldsymbol{\theta}_1$ and $\mathcal{F}(\Phi_y, p, r) = (\Phi_y^{\otimes p} \otimes \mathbb{1}^{\otimes r})$.
    Clearly, $d=2^{p+r}$ states $\ket{j}=\{\ket{i_1i_2\ldots i_{p+r}}\}$ with the entries $i_n=\{0, 1\}$ is a span of the Hilbert space of the registers $P$ and $R$.
    In our numerical simulations, we split the computational basis in two parts of $d/2$ basis vectors each and associate the measurement outcomes with the projectors
    \begin{equation} \label{eq:pi_projectors}
        \Pi_0 = \sum_{j=1}^{d/2} \ketbra{j}{j}, \quad \Pi_1 = \sum_{j=d/2+1}^{d}\ketbra{j}{j}.
    \end{equation}
    As the result, we come up with the optimization problem that can be addressed using a hybrid quantum-classical setup:
    \begin{equation} \label{eq:psuc_par_var}
        \mathrm{p_s^{par}} = \frac{1}{2} \max_{\boldsymbol{\theta}, r}\{\Tr(\Pi_0\, \rho^\mathrm{out}(\boldsymbol{\theta}, \Phi_0, p, r))  + \Tr(\Pi_1\, \rho^\mathrm{out}(\boldsymbol{\theta}, \Phi_1, p, r))\}.
    \end{equation}
    Thus, \eqref{eq:psuc_par_var} together with the circuit in Fig.~\ref{fig:par_circ_var} define the framework for numerically testing the parallel discrimination strategy.
    
    \subsubsection{Sequential strategy}

\begin{figure}[h]
        \centering
        \mbox{
            $$
            \Qcircuit @C=1.0em @R=1.0em {
            & \ket{0}             & & \qw & \ustick{P}\qw & \qw & \multigate{2}{U(\boldsymbol{\theta}_0)} & \gate{\Phi_y} & \multigate{2}{U(\boldsymbol{\theta}_1)} & \gate{\Phi_y} & \qw & \qw & \qw    & \qw & \multigate{2}{U(\boldsymbol{\theta}_{p-1})} & \gate{\Phi_y} & \multigate{2}{U(\boldsymbol{\theta}_{p})} & \meter \\
            &                     & &     &               &     &                                         &               &                                         &               &     &     & \cdots &     &                                             &               &                                             & \\
            & \ket{0}^{\otimes r} & & \qw & \dstick{R}\qw & \qw & \ghost{U(\boldsymbol{\theta}_0)}        & \qw           & \ghost{U(\boldsymbol{\theta}_1)}        & \qw           & \qw & \qw & \qw    & \qw & \ghost{U(\boldsymbol{\theta}_{p-1})}        & \qw           & \ghost{U(\boldsymbol{\theta}_{p})}        & \meter
            }
            $$
        }
        \caption{A variational quantum circuit implementing the sequential channel discrimination strategy from Fig.~\ref{fig:sequential_schem}.
        The Alice's channels $\mathcal{E} = \{ \varepsilon_j \}_{j=1}^{p-1}$ are replaced by the parametrized unitaries $\{U(\boldsymbol{\theta}_j) \}_{j=1}^{p-1}$, where $p$ is the number of allowed applications of the channel $\Phi_y$.
        The input state $\rho_{PR}^\mathrm{in}$ is prepared from $\ket{0}_P \otimes \ket{0}^{\otimes r}_R$ via the unitary transformation $U(\boldsymbol{\theta}_0)$, while $U(\boldsymbol{\theta}_p)$ is used to rotate the measurement basis.
        This method necessitates $(1+r)$ qubits, with one qubit in the register $P$ and $r$ qubits in the register $R$.
        In analogy with the parallel strategy, $r$ might be set to zero.}
        \label{fig:seq_circ_var}
    \end{figure}
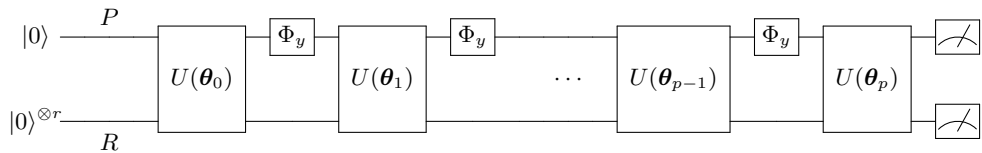

    Similarly to the parallel channel discrimination strategy, in Fig.~\ref{fig:seq_circ_var} we depict the sequential strategy \eqref{eq:p_suc_seq} formulated in the framework of variational circuits.
    In this variational approach, the Alice's channels $\mathcal{E}$ can be applied to the input state $\rho_{PR}^\mathrm{in}$ via the Stinespring representation~\cite{biamonte2019lectures}.
    Particularly, $\varepsilon[\rho_{PR}]$ can be implemented by adding an ancillary register $E$ in the state $\rho_E$ of $e$ qubits and performing a unitary evolution $U$ of the joint state $\rho_{PR} \otimes \rho_{E}$, followed by tracing out the register $E$. That is, we have
    \begin{equation} \label{eq:stinespring}
        \varepsilon[\rho_{PR}] = \Tr_E \big[ U (\rho_{PR} \otimes \rho_E \big) U^\dagger \big].
    \end{equation}
    For this transformation to be general, the register $E$ should contain $e$ qubits twice as the size of the registers $P$ and $R$ together \cite{nielsen2002quantum}.
    In our setting, this is equal to $e=2(1+r)$ with one qubit in $P$ and $r$ qubits in $R$.
    
    However, in the follow up analysis, we reduce the transformation~\eqref{eq:stinespring} to the one shown in Fig.~\ref{fig:seq_circ_var}.
    In this approach, we incorporate the register $E$ into $R$.
    Furthermore, instead of the channels $\mathcal{E} = \{ \varepsilon_j \}_{j=1}^{p-1}$ we use parametrized unitaries $\{U(\boldsymbol{\theta}_j) \}_{j=1}^{p-1}$, alongside with the operator $U(\boldsymbol{\theta}_{0})$ which prepares the initial state and the operator $U(\boldsymbol{\theta}_{p})$ which rotates the measurement basis, as done in~\eqref{eq:out_state_par_var}.
    Analogous to \eqref{eq:out_state_seq}, the output state is
    \begin{equation}
        \label{eq:out_state_seq_var}
        \rho^\mathrm{out}\big(\boldsymbol{\theta}, \Phi_y, p, r \big) = \mathcal{C}\big(\boldsymbol{\theta}, \Phi_y, p, r\big) [\rho_0(r)],
    \end{equation}
    where $\rho_0(r) = \ketbra{0}{0}_P \otimes \ketbra{0}{0}^{\otimes r}_R$ and
    \begin{equation*}
        \mathcal{C}\big(\boldsymbol{\theta}, \Phi_y, p, r \big) = \mathcal{U}(\boldsymbol{\theta}_{p}) \circ 
        (\Phi_y\otimes \mathbb{1}^{\otimes r}) \circ 
        \mathcal{U}(\boldsymbol{\theta}_{p-1}) \circ 
        \cdots \circ 
        (\Phi_y\otimes \mathbb{1}^{\otimes r}) \circ
        \mathcal{U}(\boldsymbol{\theta}_{1}) \circ 
        (\Phi_y\otimes \mathbb{1}^{\otimes r}) \circ
        \mathcal{U}(\boldsymbol{\theta}_{0}),
    \end{equation*}
    with $\boldsymbol{\theta} = \bigcup_{k=0}^{p} \boldsymbol{\theta}_k$. The channels $\mathcal{U}$ should be interpreted as in \eqref{eq:out_state_par_var}. 
    Note that there is also $U(\boldsymbol{\theta}_0)$ which is used to prepare the input state $\rho_{PR}^\mathrm{in} = \mathcal{U}(\boldsymbol{\theta}_0)[\rho_0(r)]$. 
    Technically, in line with Fig.~\ref{fig:seq_circ_var}, this can be viewed as a channel $\varepsilon_0$ which maps the single-qubit state $\ketbra{0}{0}_P$ from the Hilbert space $\mathcal{H}_P$ to a $(1+r)$-qubit state in $\mathcal{H}_P \otimes \mathcal{H}_R$ as
    \begin{equation}
        \label{eq:single-to-many-channel}
        \rho_{PR}^\mathrm{in} = \varepsilon_0[\ketbra{0}{0}_P] = E_0\ketbra{0}{0}_P E_0^\dagger + E_1\ketbra{0}{0}_P E_1^\dagger,
    \end{equation}
    where the Kraus operators are $E_j = U(\boldsymbol{\theta}_0) \big(\ket{j}_P \otimes \ket{0}_R\big) \bra{j}_P$. The rest of the transformations are unitary and performed in the extended Hilbert space $\mathcal{H}_P \otimes \mathcal{H}_R$.
    
    The output state \eqref{eq:out_state_seq_var} is then measured using the POVM elements \eqref{eq:pi_projectors} which leads to the optimization problem
    \begin{equation} \label{eq:psuc_seq_var}
        \mathrm{p_s^{seq}} = \frac{1}{2} \max_{\boldsymbol{\theta}, r} \big\{ \Tr(\Pi_0\,\rho^\mathrm{out}(\boldsymbol{\theta}, \Phi_0, p, r)) + \Tr(\Pi_1\, \rho^\mathrm{out}(\boldsymbol{\theta}, \Phi_1, p, r)) \big\}.
    \end{equation}
    As one can notice, the expressions for $\mathrm{p_s^{par}}$ in \eqref{eq:psuc_par_var} and $\mathrm{p_s^{seq}}$ in \eqref{eq:psuc_seq_var} do not differ much except for the structure and the genesis of the output state $\rho^\mathrm{out}$.
    In variational quantum computing, one may consider these expressions as objective functions maximization of which leads to training of the corresponding circuits.
    
\subsection{Numerical experiments}

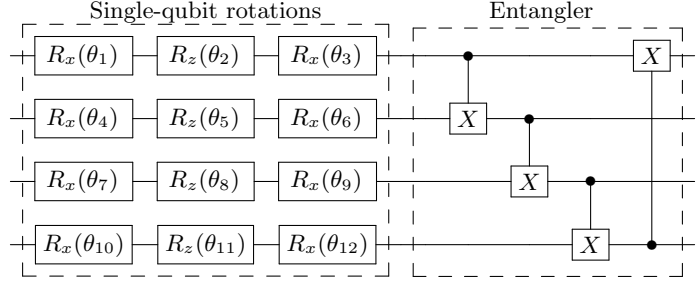
\begin{figure*}[h]
        \mbox{
            \Qcircuit @C=1.em @R=1.em {
                & & \mbox{Single-qubit rotations} & & & & & \mbox{\quad Entangler} & & & \\
                & \gate{\,R_x(\theta_{1})\;} & \gate{\,R_z(\theta_{2})\;} & \gate{\,R_x(\theta_{3})\;} & \qw & \qw & \ctrl{1} & \qw & \qw & \gate{X} & \qw \\
                & \gate{\,R_x(\theta_{4})\;} & \gate{\,R_z(\theta_{5})\;} & \gate{\,R_x(\theta_{6})\;} & \qw & \qw & \gate{X} & \ctrl{1} & \qw & \qw & \qw \\
                & \gate{\,R_x(\theta_{7})\;} & \gate{\,R_z(\theta_{8})\;} & \gate{\,R_x(\theta_{9})\;} & \qw & \qw & \qw & \gate{X} & \ctrl{1} & \qw & \qw \\
                & \gate{R_x(\theta_{10})} & \gate{R_z(\theta_{11})} & \gate{R_x(\theta_{12})} & \qw & \push{\rule[0em]{0em}{1.6em}}\qw & \qw & \qw & \gate{X} & \ctrl{-3} & \qw \gategroup{2}{6}{5}{10}{1.em}{--}
                \gategroup{2}{2}{5}{4}{1.em}{--} 
            }   
        }
        \caption{A layer of the hardware-efficient ansatz with 4 input qubits and 12 variational parameters. Here, $R_{\sigma}(\theta) = e^{-\imath \theta\sigma}$ with $\sigma\in \{X, Y, Z\}$ specifying the Pauli operator and $\theta \in [0, 2\pi)$ being the optimization parameters.}
        \label{fig:hardware_efficient_ansatz}
    \end{figure*}

    We herein describe the results of our numerical simulations demonstrating the capability of variational quantum channel discrimination. Although the number of repetitions $p$ and the number of qubits $r$ in the register $R$ enter the equations \eqref{eq:psuc_par_var} and \eqref{eq:psuc_seq_var}, we do not optimize with respect to these parameters. 
    Instead, we fix $p=1,2$ and vary the amount of ancillary qubits $r$ upon optimizing the ansatz parameters $\boldsymbol{\theta}$.
    The parametrized unitary operators $U(\boldsymbol{\theta}_k)$ are implemented in terms of the hardware-efficient ansatz~\cite{kandala_hardware-efficient_2017} whose four-qubit structure is shown in Fig.~\ref{fig:hardware_efficient_ansatz}.
    This circuit is composed of several layers constituted by universal the single-qubit rotations and an entanglement block.
    In this ansatz, the number of variational parameters $s$ is polynomial in the total number of qubits $q=p+r$ and the number of layers $l$, $s = 3ql$ for $q > 2$.
    In what follows, we test the variational approach to discriminate the depolarizing channels and entanglement breaking channels, mapping two-qubit states into one-qubit states.
    
\subsubsection{Entanglement breaking channel discrimination}

    We start our numerical analysis with the variational circuit approach to discriminating the entanglement breaking channels 
    \begin{equation} \label{eq:entbreak_ch}
        \Phi_0[\rho] = \sum_{j=1}^5 A_j \rho A_j^\dagger, \quad
        \Phi_1[\rho] = \sum_{j=1}^5 B_j \rho B_j^\dagger,
    \end{equation}
    described by the Kraus operators
    \begin{align}
        A_1 = \ketbra{0}{00},\; A_2 = \ketbra{0}{01},\; A_3 = \ketbra{0}{10},\; A_4 = \frac{1}{\sqrt{2}}\ketbra{0}{11},\; A_5 = \frac{1}{\sqrt{2}}\ketbra{1}{11}, \label{eq:a_ops}\\
        B_1 = \ketbra{+}{00},\; B_2 = \ketbra{+}{01},\; B_3 = \ketbra{1}{1+},\; B_4 = \frac{1}{\sqrt{2}}\ketbra{0}{1-},\; B_5 = \frac{1}{\sqrt{2}}\ketbra{1}{1-} \label{eq:b_ops}
    \end{align}
    with $\ket{\pm} = (\ket{0} \pm \ket{1}) / \sqrt{2}$.
    In this particular scenario, the parallel discrimination strategy was demonstrated to never reach unity of the successful discrimination probability \eqref{eq:p_suc_diamond_par}, i.e. $\mathrm{p_\diamond^{par}}(p) < 1$  for any finite number of channel applications $p$ \cite{harrow2010adaptive}.
    At the same time, the sequential strategy with only $p=2$ repetitions allows one to distinguish the channels perfectly, and the input state does not need to be entangled.
    To estimate $\mathrm{p_\diamond^{par}}$ defined in  \eqref{eq:p_suc_diamond_par}, one needs to calculate the diamond norm which can be done via semi-definite programming~\cite{watrous2018theory}.
    Using the CVXPY package~\cite{agrawal2018rewriting, diamond2016cvxpy}, we calculated the probability $\mathrm{p_\diamond^{par}}(p)$ for $p=1, 2$; our results reveal that $\mathrm{p_\diamond^{par}}\approx 0.9268$ for $p=1$ and $\mathrm{p_\diamond^{par}}\approx 0.9771$ for $p=2$.
    
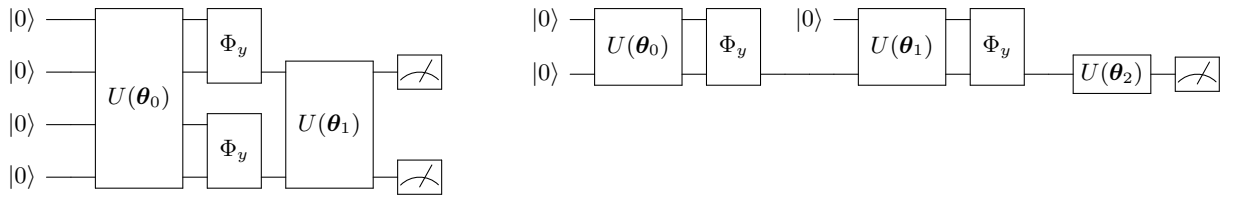
\begin{figure}[h]
        \centering
        \mbox{
            \Qcircuit @C=1.0em @R=1.0em {
                & \ket{0} & & \qw    & \multigate{3}{U(\boldsymbol{\theta}_0)} & \multigate{1}{\Phi_y} &                  & \\
                & \ket{0} & & \qw    & \ghost{U(\boldsymbol{\theta}_0)}        & \ghost{\Phi_y}        & \multigate{2}{U(\boldsymbol{\theta}_1)} & \meter \\
                & \ket{0} & & \qw    & \ghost{U(\boldsymbol{\theta}_0)}        & \multigate{1}{\Phi_y} &                  & \\
                & \ket{0} & & \qw    & \ghost{U(\boldsymbol{\theta}_0)}        & \ghost{\Phi_y}        & \ghost{U(\boldsymbol{\theta}_1)}        & \meter \\
            }
        }
        \hfil
        \mbox{
            $$
            \Qcircuit @C=1.0em @R=1.0em {
                & \ket{0} & & \multigate{1}{U(\boldsymbol{\theta}_0)} & \multigate{1}{\Phi_y} &     & \ket{0} &     & \multigate{1}{U(\boldsymbol{\theta}_1)} &  \multigate{1}{\Phi_y} &     &                                         & \\
                & \ket{0} & & \ghost{U(\boldsymbol{\theta}_0)}        & \ghost{\Phi_y}        & \qw & \qw     & \qw & \ghost{U(\boldsymbol{\theta}_1)}        &  \ghost{\Phi_y}        & \qw &  \gate{U(\boldsymbol{\theta}_2)}        & \meter \\
            }
            $$
        }
        \caption{The explicit variational quantum circuits trained for discriminating the entanglement breaking channels \eqref{eq:entbreak_ch} with $p=2$ channel applications. 
        On the left shown is the circuit for the parallel strategy and on the right is for the sequential strategy. 
        Note that the channels to be discriminated map two-qubit states into one-qubit states.
        Therefore for the sequential strategy, after the first application of the channel $\Phi_y$, one has to add an extra qubit in some state (in our case, $\ket{0}$).}
        \label{fig:ent_br_circs}
    \end{figure}

    We proceed by training the variational circuits depicted in Figs.~\ref{fig:par_circ_var},\ref{fig:seq_circ_var}.
    Recall that the circuits are trained by maximizing \eqref{eq:psuc_par_var} and \eqref{eq:psuc_seq_var}, respectively. For this task, we made use of the L-BFGS-B \cite{byrd1995limited} optimization method.
    The explicit quantum circuits to be trained to discriminate the entanglement breaking channels with $p=2$ repetitions are shown in Fig.~\ref{fig:ent_br_circs} for both the parallel and sequential strategies. 
    In the parallel strategy, for $p=2$ the success probability $\mathrm{p_s^\mathrm{par}}\approx \mathrm{p_\diamond^{par}}\approx 0.9771$ was achieved with the use of the hardware-efficient ansatz of $l=5$ layers representing $U(\boldsymbol{\theta}_{0})$ and $U(\boldsymbol{\theta}_{1})$. 
    Note that to get this probability no ancillary qubits are needed, i.e. $r=0$.
    In the sequential strategy, even a one-layer, $l=1$, hardware-efficient ansatz parametrizing the operators $U(\boldsymbol{\theta}_0)$, $U(\boldsymbol{\theta}_1)$ and $U(\boldsymbol{\theta}_2)$ provides the success probability $\mathrm{p_s^{seq}} \approx 1$ for $p=2$.
    Note that for $p=1$, the circuits for both parallel and sequential discrimination look alike, and using a one-layer hardware-efficient ansatz results in $\mathrm{p_s}\approx \mathrm{p}_\diamond \approx 0.9268$.
    
    It however might be excessive to use circuits with that many parameters, especially in the case of the sequential strategy. A successful discrimination, as implemented in the seminal work~\cite{harrow2010adaptive}, did not use anything but the channels $\Phi_y$ and a single measurement at all. 
    Meanwhile, even with these (over)parametrized circuits one is capable of identifying optimal input states and measurement bases such that $\mathrm{p_s^{seq}} \approx 1$.
    It is thus reasonable to expect that the variational approach could be useful for channels $\Phi_y$ without any prior knowledge of them.
    
\subsubsection{Depolarizing channel discrimination}
    
    Our study is continued with training the variational quantum circuits for discriminating depolarizing channels as given by
    \begin{equation}
        \label{eq:dep_channel}
        \Phi(\alpha)[\rho] = (1 - \alpha) \rho + \frac{\alpha}{3}\big(\sigma_x \rho \sigma_x + \sigma_y \rho \sigma_y + \sigma_z \rho \sigma_z \big)
    \end{equation}
    where the coefficient $\alpha$ determines the depolarization factor, while $\sigma_i$ with $i=x,y,z$ stands for the Pauli operators. 
    We consider a pair of channels with $0\leqslant \alpha_0\neq\alpha_1\leqslant 1$. In our numerical simulations, we fixed the number of channel applications to $p=2$; the number of qubits in the ancillary register $R$ is set to $r=3$ for the parallel strategy and $r=4$ for the sequential strategy, so that the total number of qubits was $q=5$ for both cases (see Fig.~\ref{fig:par_circ_var} and Fig.~\ref{fig:seq_circ_var}).
    
\begin{figure*}[h]
        \centering
        \includegraphics[width=.475\textwidth]{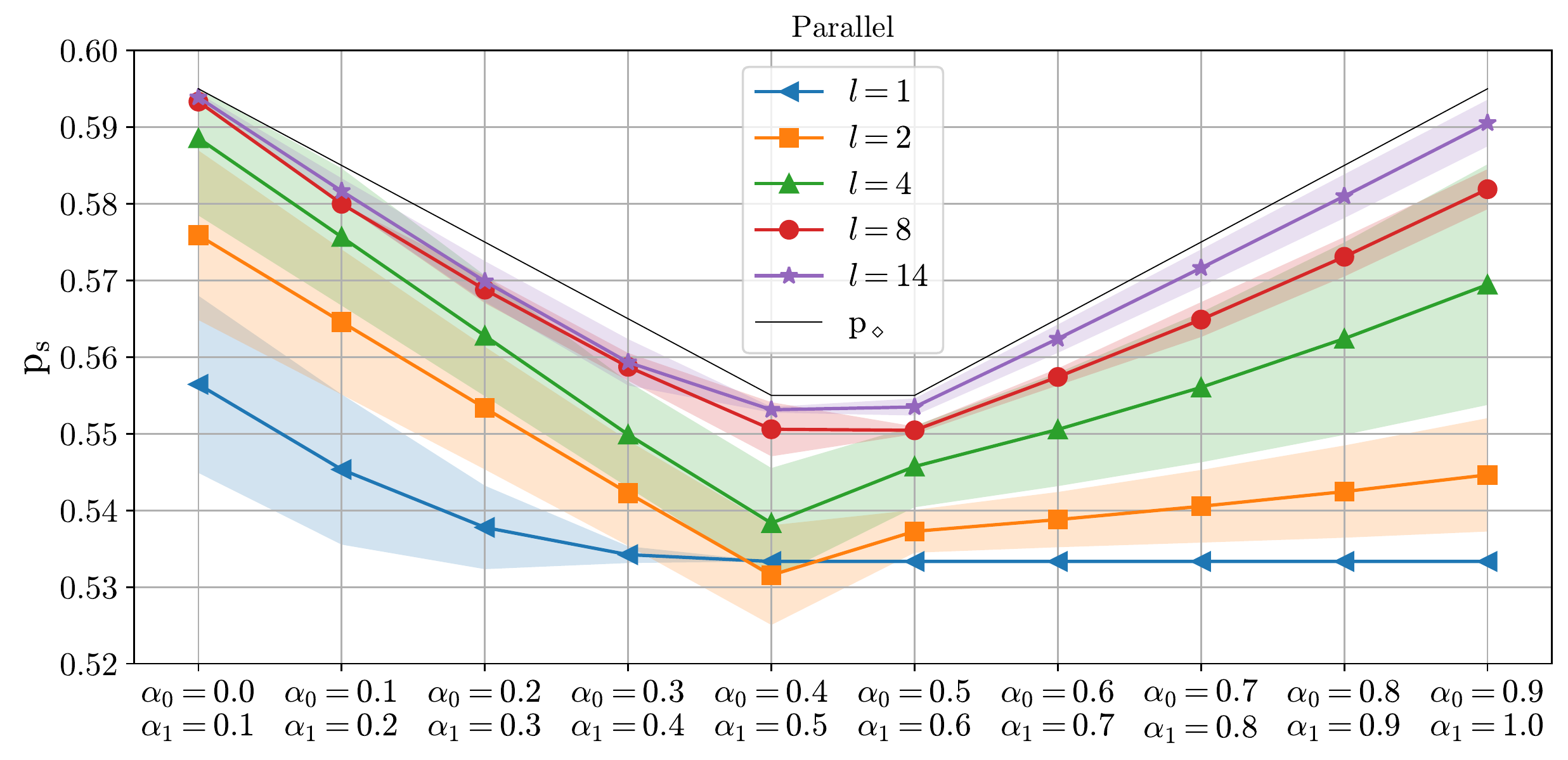}
        \includegraphics[width=.475\textwidth]{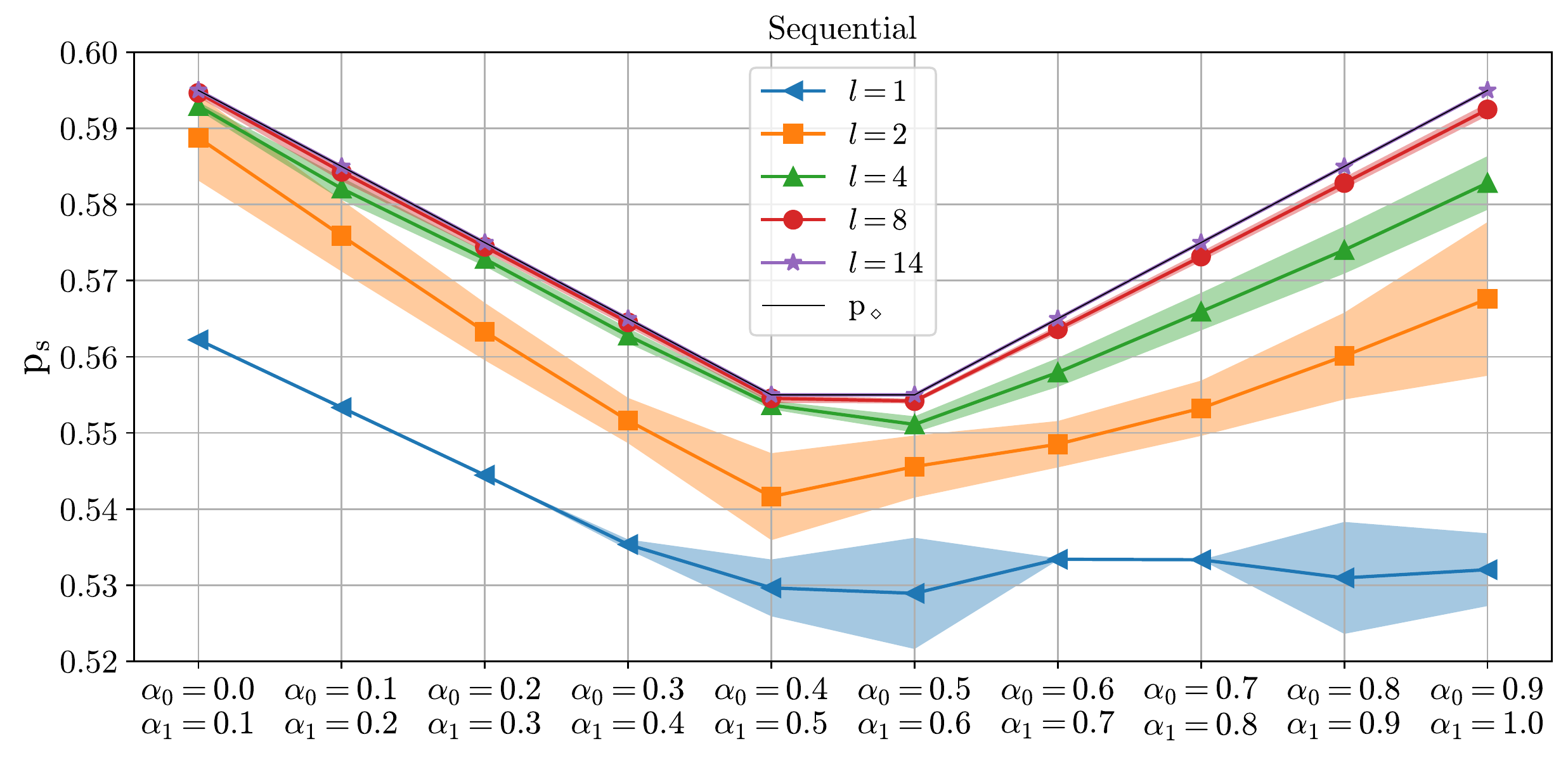}
        \caption{Probability of successful discrimination $\mathrm{p_{suc}}$ between a pair of depolarizing channels with $\alpha_0$ and $\alpha_1$ achieved with the parallel (left panel) and sequential (right panel) strategy. With different colors shown are the results for specific number of ansatz layers $l$. Marks connected by solid lines stand for the average probabilities obtained after ten independent runs, and shaded areas are to show the standard deviation. The black solid line indicates the maximum achievable probability for the parallel strategy $\mathrm{p_\diamond^{par}}$ for $p=2$ channel evaluations; for the sequential strategy, this line overlaps with the purple curve corresponding to $l=14$.}
        \label{fig:psucs-par-seq}
    \end{figure*}

    In Fig.~\ref{fig:psucs-par-seq} shown are the probabilities \eqref{eq:psuc_par_var} and \eqref{eq:psuc_seq_var} for both parallel and sequential strategy, respectively, as evaluated for a pair $(\alpha_0, \alpha_1 = \alpha_0 + 0.1)$ starting from $\alpha_0=0$. To speed up the calculations, one might use random ansatz parameters for the initial guess at $(\alpha_0=0.0, \alpha_1 = 0.1)$; whereas for each next pair up to $(\alpha_0=0.4, \alpha_1 = 0.5)$ the optimal parameters are taken as obtained at an earlier step. The same strategy can be successfully adopted for a set of parameters starting from $(\alpha_0=0.9, \alpha_1 = 1.0)$ with random initialization $\boldsymbol{\theta}$ down to $(\alpha_0=0.5, \alpha_1 = 0.6)$.
    
    We have also explored how the achieved probabilities $\mathrm{p_{s}}$ depend on $l$, the number of layers of hardware-efficient ansatz in variational circuits.
    Clearly, the sequential strategy gives better results and smaller variance with fewer number of layers.
    In this strategy, $l=14$ layers turns out to be enough for achieving the success probability $\mathrm{p_\diamond^{par}}$ for all pairs $(\alpha_0, \alpha_1)$. 
    In the parallel strategy, this result cannot be reproduced no matter how big the value $l$ is (we tested for up to $l=30$ layers). 
    One may also notice that despite the same diamond-distances~\eqref{eq:p_suc_diamond_par}, it is harder to achieve higher $\mathrm{p_\diamond^{par}}$ for depolarization factor pairs on the right to $\alpha=0.5$.
    As an instance, for the pairs $(\alpha_0=0.0, \alpha_1=0.1)$ and $(\alpha_0=0.9, \alpha_1=1.0)$ the theoretical probabilities are $\mathrm{p_\diamond^{par}} = 0.595$, but the obtained probabilities $\mathrm{p_{s}}$ are lower for the second pair of depolarization factors.
    We conclude that there might be some dependence on the trace product between the states passed through the channels, $\Tr(\rho_0 \rho_1)$ with $\rho_y \equiv \Phi(\alpha_y)[\rho]$ (see Appendix \ref{app:trace_prod} for details).
    
\section{Binary quantum classifier}

    Previously, we discussed how one can use variational quantum computing framework to reformulate and solve the problem of channel discrimination.
    With a proper post-processing of measurements, variational quantum circuits can also serve as a means to solve classification tasks.
    For example, one might think of classifying phases of matter~\cite{Shirinyan_2019,uvarov2020machine,Berezutskii_2020}.
    That is, having a variational classifier trained on labeled data points (quantum states of different phases), one may predict unknown labels of given states.
    In this section, we solve a similar problem, limiting ourselves to binary classification of the depolarizing channel \eqref{eq:dep_channel} with two different depolarization factors $\alpha_0$ and $\alpha_1$.
    
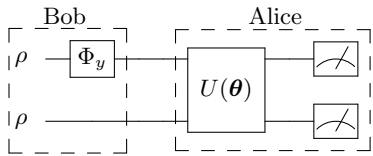
\begin{figure*}[h]
        $$\Qcircuit @C=1.0em @R=1.0em {
           & & \;\;\quad\mbox{\text{Bob}} & & & & \qquad\qquad\mbox{\text{Alice}} & & \\
           & \rho & & \gate{\Phi_y}                    & \qw & \qw & \multigate{1}{U(\boldsymbol{\theta})} & \qw & \meter \\
           & \rho & & \push{\rule[0em]{0em}{1.6em}}\qw & \qw & \qw & \ghost{U(\boldsymbol{\theta})}        & \qw & \meter\gategroup{2}{2}{3}{4}{1.em}{--}\gategroup{2}{7}{3}{9}{1.em}{--}\\
        }$$
        \caption{A schematic of the variational classifier of quantum channels. 
        Bob prepares the states $\Phi_y[\rho] \otimes \rho$ for Alice who then applies the unitary $U(\boldsymbol{\theta})$ and measures the resultant state in the computational basis. 
        The measurement results are further used to compute the prediction value $p$ as yielded by Eq.~\eqref{eq:label_func}.
        To train the circuit, Alice minimizes the square distances \eqref{eq:ls-loss} between the predictions $p$ and the true labels $y$.}
        \label{fig:qq-scheme}
    \end{figure*}

    Similarly to the variational channel discrimination discussed in Section \ref{sec:quantum_channel_discrimination}, the problem of building a variational classifier of quantum channels can be described in the form of a game between Alice and Bob.
    In this game, Alice wants to train a variational circuit $U(\boldsymbol{\theta})$ such that given an output state of a channel $\Phi_y$ from Bob, after post-processing the results of measurements, this circuit allows to predict the label $y$ of the channel applied. 
    There are several peculiarities in the game we consider.
    First, Bob sends to Alice not only the output state $\rho_y = \Phi_y[\rho]$, but the original state $\rho$ as well, i.e. from Bob Alice receives the states $\rho_y \otimes \rho$.
    Second, Alice does not control the original state $\rho$: it is prepared by Bob, and it is random and mixed.
    Third, the post-processing of measurement results assumes that Alice is allowed to perform arbitrarily many measurements; or, equivalently, Bob is assumed to give an arbitrary number of copies of the channel's output state and the corresponding original states.
    This game is schematically shown in Fig.~\ref{fig:qq-scheme}.

    Let us describe the game more formally.
    At the beginning, Bob picks two values $0\leqslant \alpha_0 \neq \alpha_1\leqslant 1$.
    Bob then selects a label $y \in \{0, 1\}$, creates two copies of a randomly generated (in general, mixed) qubit state $\rho$, and passes one of these states through a quantum channel giving $\rho_y = \Phi(\alpha_y)[\rho]$.
    The other copy of the state remains unaffected. Bob sends the state $\rho_y \otimes \rho$ to Alice who feeds this as an input for the variational circuit $U(\boldsymbol{\theta})$, and the resultant state is determined by
    \begin{equation}
        \label{eq:joined_state}
        \rho(\alpha_y, \boldsymbol{\theta}) = U(\boldsymbol{\theta})(\rho_y \otimes \rho)U^\dagger(\boldsymbol{\theta}).
    \end{equation}
    Alice measures the observable $\sigma_z \otimes \sigma_z$ for calculating 
    \begin{equation}
        \label{eq:label_func}
        p(\boldsymbol{\theta}) = \frac{1}{2} \big(1 + \Tr[ \rho(\alpha_y, \boldsymbol{\theta}) \, (\sigma_z \otimes \sigma_z)]\big), 
    \end{equation}
    that is used to quantify the prediction of the label $y$.
    The task of Alice is to train the circuit $U(\boldsymbol{\theta})$ such that given a pair $\rho_y \otimes \rho$ one is able to predict the label $y$ of the depolarizing channel based on \eqref{eq:label_func}, i.e. for some $0< b <1$ Alice returns $y=0$ if $p \leqslant b$, and $y=1$ otherwise.

    To train the circuit, Bob provides Alice with the training set $\{\rho_{y_j}^j \otimes \rho^j, y_j \}_{j=1}^{N_{\text{train}}}$ where $y_j \in \{0, 1\}$ are true labels and the superscript in $\rho^j$ indicates that each state is different since generated randomly. 
    Then Alice feeds each pair into the circuit $U(\boldsymbol{\theta})$ and calculates the predictions $p_j(\boldsymbol{\theta})$ as defined by \eqref{eq:label_func}.
    Having obtained $\{p_j(\boldsymbol{\theta}), y_j \}_{j=1}^{N_{\text{train}}}$, Alice makes use of the least squares loss function to determine the optimal circuit parameters $\boldsymbol{\theta}$:
    \begin{equation}
        \label{eq:ls-loss}
        f(\boldsymbol{\theta}) = \sum_{j=1}^{N_\text{train}} \big( y_j - p_j(\boldsymbol{\theta}) \big)^2.
    \end{equation}
    Knowing the true labels $y_j$ and having sorted predictions $p_j$, Alice can determine the parameter $b$ that separates two classes. Calculating can be based on maximizing the accuracy during training. The search of $b$ is carried out iteratively. At each step $t$, Alice takes the prediction $p_t$ and groups all the elements that are less than or equal to $b$ to belong to the first class $(p_1,p_2,\ldots,p_t) \in\{\textrm{`0'}\}$ and to the second class $(p_{t+1},p_{t+2},\ldots,p_{N_\mathrm{train}}) \in\{\textrm{`1'}\}$ otherwise. As a result, $b$ equals to the prediction value $p_t$, the division by which gives the best accuracy during the training.
    
    Having found the optimal circuit parameters $\boldsymbol{\theta}^\mathrm{opt}$ and the border value $b$, Alice must test the obtained classifier. 
    To do that, Alice receives from Bob a set $\{\rho_{y_j}^j \otimes \rho^j \}_{j=1}^{N_{\text{test}}}$, which does not contain the true labels $y_j$.
    As during the training, Alice feeds each state $\rho_{y_j}^j \otimes \rho^j$ from the test set to the circuit $U(\boldsymbol{\theta}^\mathrm{opt})$, computes the corresponding prediction values $p_j$ in \eqref{eq:label_func} and assigns the label $y_j=0$ if $p_j \leqslant b$ or $y_j=1$ if $p_j \geq b$
    
    To test the described approach of building a variational channel classifier, we performed numerical experiments of distinguishing the depolarizing channels \eqref{eq:dep_channel} with different depolarization factors $\alpha_0$ and $\alpha_1$.
    To represent the variational circuit $U(\boldsymbol{\theta})$, we considered the three ans\"atze:
    \begin{align}
        U_1(\boldsymbol{\theta}) &= CR_y^{12}(\theta_7) [ R_x(\theta_3) R_z(\theta_2)R_x(\theta_1)] \otimes [ R_x(\theta_6) R_z(\theta_5) R_x(\theta_4) ], \label{eq:class_u1} \\
        U_2(\boldsymbol{\theta}) &= [ R_z(\theta_2) R_x(\theta_1) ] \otimes [ R_z(\theta_4) R_x(\theta_3)], \label{eq:class_u2} \\
        U_3(\boldsymbol{\theta}) &= R_z(\theta_2) R_x(\theta_1). \label{eq:class_u3}
    \end{align}
    The first ansatz $U_1(\boldsymbol{\theta})$, where $CR_y^{12}(\theta)$ represents a controlled $Y$-rotation with the first control and the second target qubit, is up to the two-qubit gate essentially a hardware-efficient ansatz of a single layer shown in Fig.~\ref{fig:hardware_efficient_ansatz}. 
    The second ansatz $U_2(\boldsymbol{\theta})$ is a truncated realization of $U_1(\boldsymbol{\theta})$ with no entangling gate present.
    The classifiers built on the ans\"atze $U_1(\boldsymbol{\theta})$ and $U_2(\boldsymbol{\theta})$ are two-qubit, and they are trained on the pairs $\rho_{y_j} \otimes \rho^j$, as described in the beginning of the section.
    The third ansatz $U_3(\boldsymbol{\theta})$ is of single-qubit structure, and in this case the classifier is trained on the states $\rho_{y_j}$ that passed through a channel, without feeding in the original states $\rho^j$.
    
\begin{figure*}[h]
        \centering
        \includegraphics[width=.325\textwidth]{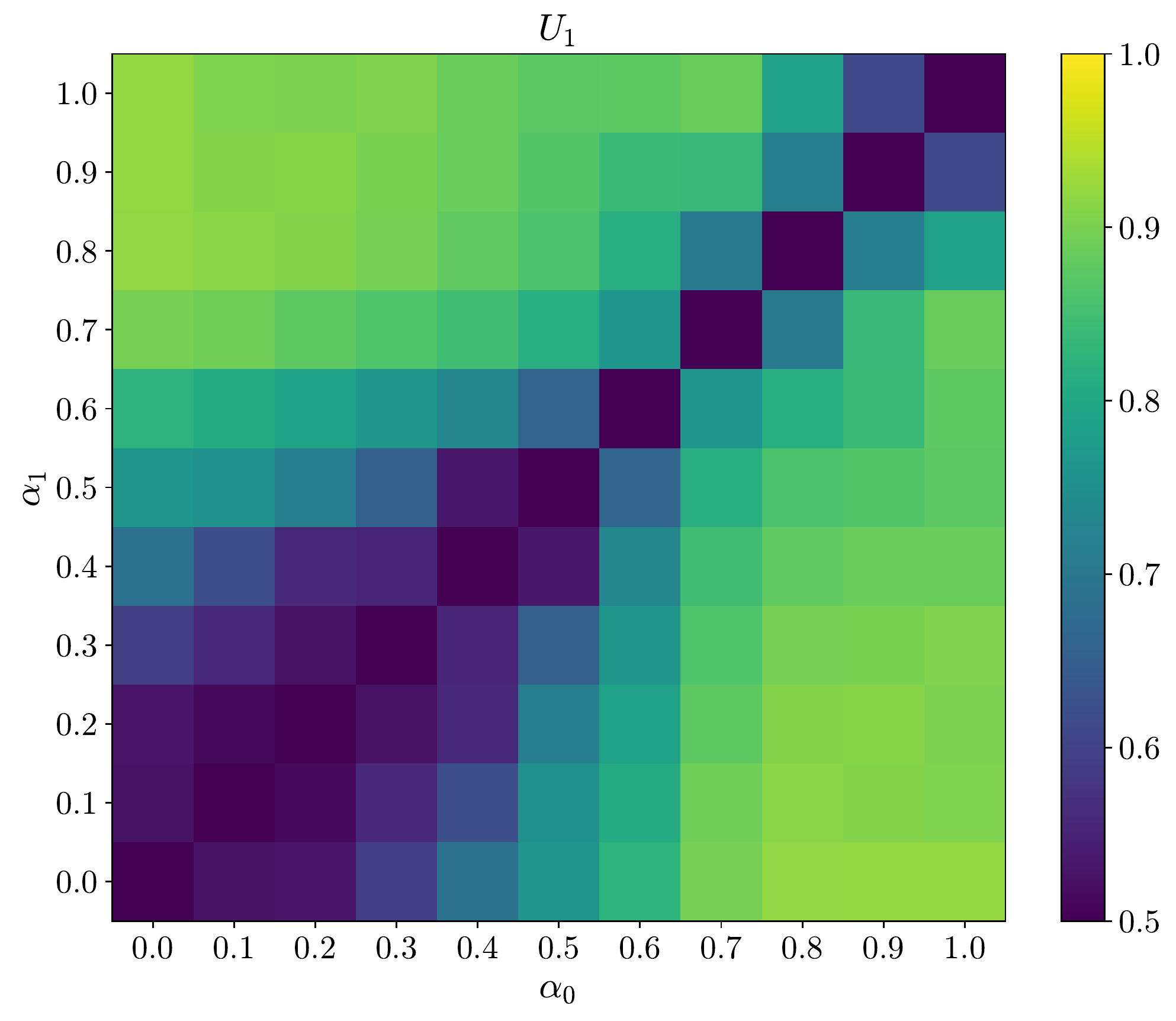}
        \includegraphics[width=.325\textwidth]{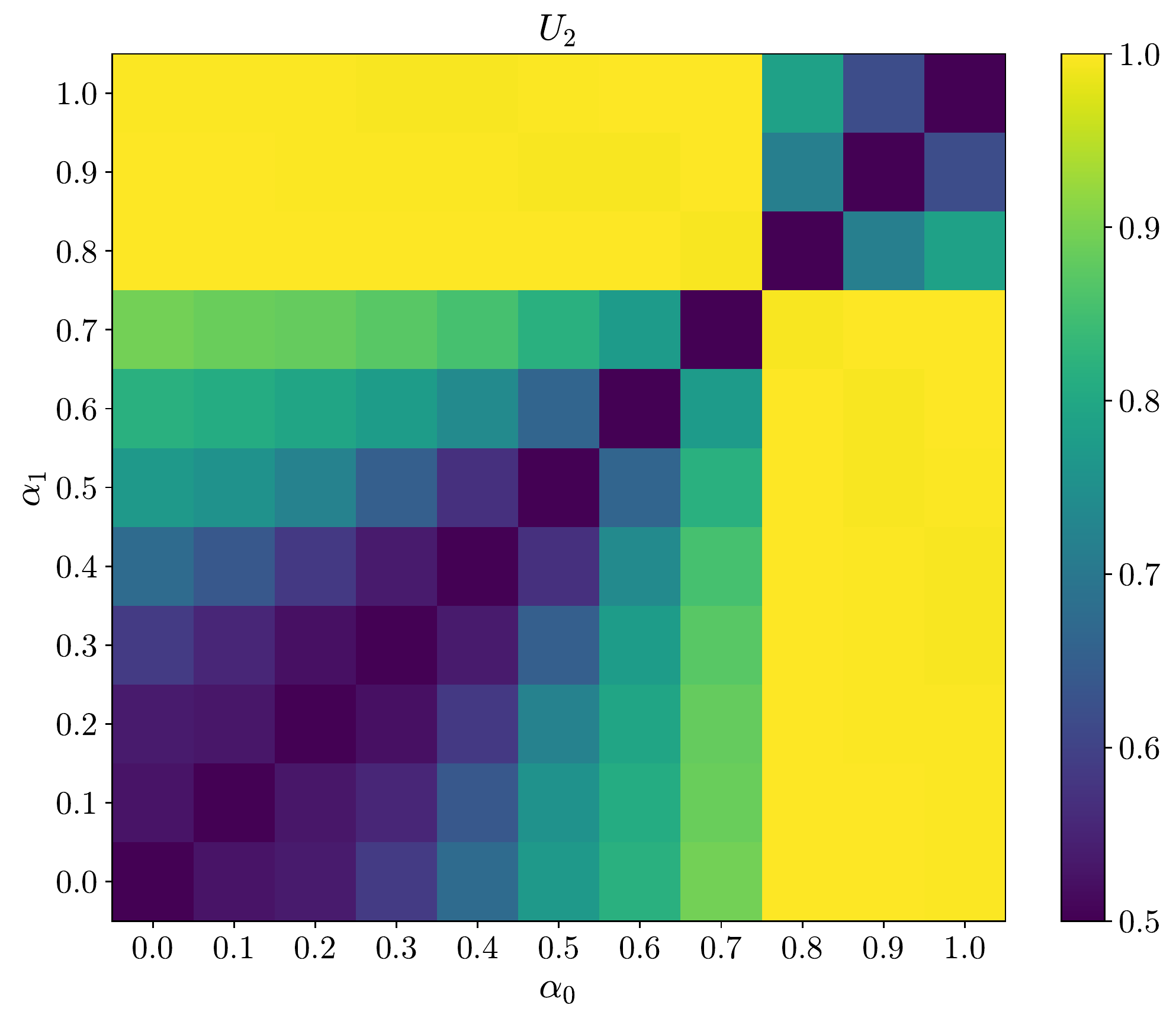}
        \includegraphics[width=.325\textwidth]{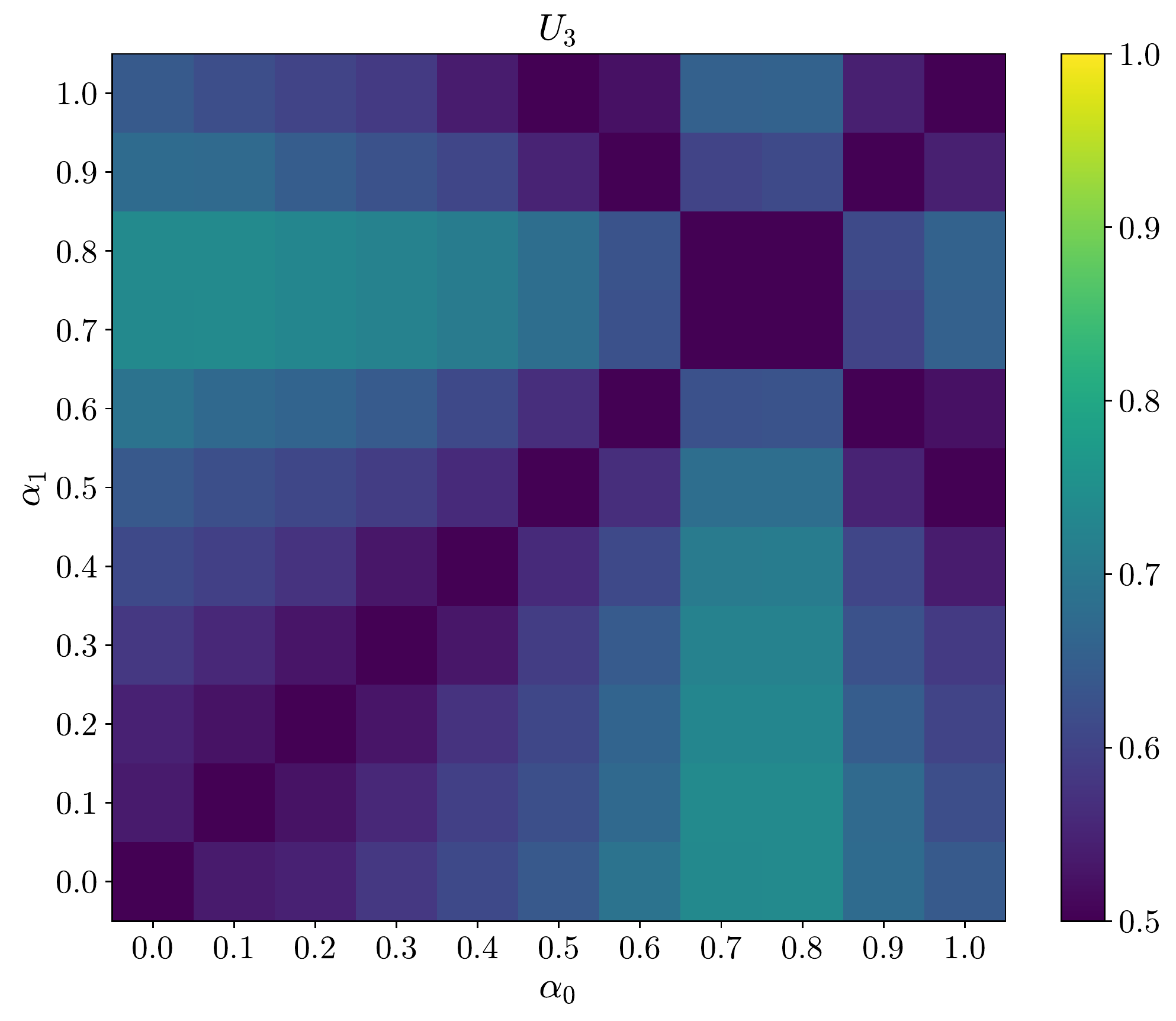}
        \caption{The accuracy of quantum channel discrimination as obtained on the test set for depolarizing channels with $\alpha_0$ and $\alpha_1$. The left, center and right panels show the accuracy of $U_1$--, $U_2$--, and $U_3$--based classifiers defined in \eqref{eq:class_u1}, \eqref{eq:class_u2} and \eqref{eq:class_u3}, respectively. The training and test sizes are $N_\text{train} = N_\text{test} = 1000$.}
        \label{fig:accuracy_heatmaps}
    \end{figure*}

    The results of our numerical experiments with the classifiers based on the ans\"atze (\ref{eq:class_u1})-(\ref{eq:class_u3}) are shown in Fig.~\ref{fig:accuracy_heatmaps}.
    These results are obtained after training the classifiers on sets of the size $N_\mathrm{train}=1000$ and tested on sets of the same size, $N_\mathrm{test}=1000$.
    A close inspection of the plots suggests that even the $U_3$--based classifier that takes only the states $\rho_{y_j}$ for training is capable of discriminating the quantum channels with some accuracy.
    Among $U_1$ and $U_2$ classifiers which are trained on pairs $\rho_{y_j} \otimes \rho^j$, the better accuracy is achieved by the one that uses a simpler though less expressive ansatz with no two-qubit gates.
    In our numerical simulations, as shown in Fig.~\ref{fig:accuracy_heatmaps} this $U_2$--based classifier unveils excellent accuracy in discriminating the channels with $\alpha_0 \lesssim 0.75 \lesssim \alpha_1$.
    Interestingly, the $U_3$--based classifier yields the highest degree of discrimination accuracy for the depolarization factors $\alpha=0.7$ or $0.8$, i.e. near $0.75$.
    This agrees with the fact that extremum of the function 
    \begin{equation*}
        \mathcal{K}(\rho_\alpha, \rho_{\alpha+\epsilon}) = \Tr(\rho_\alpha \rho_{\alpha+\epsilon}),
    \end{equation*}
    is reached at $\alpha = 0.75 - \epsilon/2$ for any $\rho \neq \mathbb{1}/2$, provided $0 \leqslant (\alpha+\epsilon) \leqslant 1$.
    This can be established by solving the equation $\partial_\alpha \Tr(\rho_\alpha \rho_{\alpha+\epsilon}) = 0$ for $\alpha$.
    
\section{Kernel-based classifier}

    An alternative way for discriminating quantum channels using a quantum processor can be traced back to the kernel methods which can be formulated as follows.
    Suppose we have a set of states $\mathcal{X} = \{\rho_i\}$. The kernel is essentially a function $\mathcal{K}:\mathcal{X} \times \mathcal{X} \rightarrow \mathbb{R}$ that guarantees the Gram matrix $\mathcal{K}_{ij} =\mathcal{K}(\rho_i, \rho_j)$ to be positive semi-definite \cite{schuld2019quantum}. In particular, the trace of the product of density operators, 
    \begin{equation}
        \label{eq:kernel}
        \mathcal{K}(\rho_i, \rho_j) = \Tr(\rho_i \rho_j),
    \end{equation}
    mentioned in previous sections, does possess such properties. The kernel-based classification methods are built on the so-called representer theorem \cite{scholkopf2002learning}. One can think of supervised machine learning based on the support vector machine where the so-called kernel trick is widely utilized~\cite{steinwart2008support, rahimi2007random}. 
    In this method, given a training set $\{\rho_j, y_j \}_{j=1}^{N_{\text{train}}}$ with labels $y_j \in \{-1, +1\}$, the cost function for maximization is
    \begin{equation}
        \label{eq:kernel_cost}
        f(\boldsymbol{\theta}) = \sum_{i=1}^{N_\text{train}} \theta_i - \frac{1}{2} \sum_{i,j=1}^{N_\text{train}} \theta_i\theta_j\,\mathcal{K}(\rho_i, \rho_j)\,y_iy_j
    \end{equation}
    with respect to $\boldsymbol{\theta}= \{\theta_j \}_{j=1}^{N_\mathrm{train}}$, on condition that $\sum_{i=1}^{N_\text{train}} \theta_i y_i = 0 $ and $\theta_i \geqslant 0$~\cite{havlivcek2019supervised}.
    Having found the optimal parameters $\boldsymbol{\theta}^\mathrm{opt} = \arg\max_{\boldsymbol{\theta}} f(\boldsymbol{\theta})$, one returns the labels based on the prediction function
    \begin{equation}
        \label{eq:prediction_kernel}
        p(\rho) = \sum_{i=1}^{N_\text{train}} \theta_i^\mathrm{opt} y_i\,\mathcal{K}(\rho_i, \rho) + b,
    \end{equation}
    where the bias $b$ is defined by
    \begin{equation*}
        b = \sum_{i=1}^{N_\text{train}} \theta_i^\mathrm{opt} y_i \, \mathcal{K}(\rho_i, \rho_m) - y_m
    \end{equation*}
    for any $m$ such that $\theta_m^\mathrm{opt} > 0$. In binary classification, the class ones assigns to a given $\rho$ is determined by $y=\mathrm{sgn}[p(\rho)]$.
    
    To formulate the problem of channel discrimination based on quantum kernel estimation, we again consider the game between Alice and Bob.
    Again, in this game Alice tries to discriminate two depolarizing channels of the form \eqref{eq:dep_channel}.
    However, this time the problem for Alice is harder: the channels will be associated not with fixed values of the depolarizing factors, but with ranges of it.
    
    Let us formalize the game.
    First, Alice chooses an input state $\rho^\mathrm{in}$ and sends $N_\mathrm{train}$ copies of it to Bob. Then Bob selects two intervals $\overline{\alpha}_{-1}$ and $\overline{\alpha}_{+1}$ such that $\overline{\alpha}_y \subset [0, 1]$. After that, Bob tosses a fair coin and attributes heads to $y=-1$ and tails to $y=+1$. Finally, Bob picks up a random $\alpha_{y} \in \overline{\alpha}_y$ and applies the depolarizing channel to one of the Alice's states, $\rho_{y} = \Phi(\alpha_{y})[\rho^\mathrm{in}]$.
    As was mentioned, the class labels $y=\pm 1$ are attributed not to the specific values of the depolarization factor $\alpha$, but to the fixed intervals of it.
    Having done that for all the states, Bob sends the training set $\{\rho_{y_j}^j, y_j\}_{j=1}^{N_{\text{train}}}$ to Alice who trains the classifier by maximizing the function~\eqref{eq:kernel_cost}.
    Here in $\rho_{y_j}^j$, the subscript $y_j\in\{\pm 1\}$ tells the interval the depolarization factor $\alpha$ is taken from, while the superscript $j$ highlights that this factor is in general different for different input states $\rho^\mathrm{in}$ (recall that Bob picks $\alpha_{y} \in \overline{\alpha}_y$ randomly). 
    Note that Bob does not tell Alice the intervals $\overline{\alpha}_y$ or the depolarizing factors $\alpha_y$ that were chosen, Alice knows only their true labels $y \in \{-1, +1\}$. 
    To test the classifier, Alice sends $N_\mathrm{test}$ copies of the state $\rho^\mathrm{in}$ to Bob, who sends back the test set $\{\rho_{y_j}^j = \Phi(\alpha^j_{y_j})[\rho^\mathrm{in}] \}_{j=1}^{N_{\text{test}}}$ prepared similarly to the training one. 
    For each state $\rho_j \equiv \rho_{y_j}^j$ of the test set, Alice calculates the prediction $p(\rho_j)$ as specified by \eqref{eq:prediction_kernel} and assigns to this prediction the class label $y_j = \mathrm{sgn}[p(\rho_j)]$.
    In practice, Alice could estimate the kernel $\mathcal{K}(\rho_i, \rho_j) = \Tr(\rho_i\rho_j)$ via the so-called controlled-SWAP test routine \cite{kobayashi2003quantum}. 
    
    To test the kernel-based approach of classification, we performed numerical experiments by training such a classifier on sets of the size $N_{\text{train}} = 100$ and testing it on sets of the size $N_{\text{test}} = 1000$.
    In what follows, we consider four classifier instances trained for discriminating the channels with the following pairs of intervals of the depolarization factors $\alpha$: 
    \begin{align}
        I_1 &= \big\{ \overline{\alpha}_{-1}=[0.0, 0.5),\; \overline{\alpha}_{+1}=[0.5, 1.0] \big\},   \label{eq:int_1} \\
        I_2 &= \big\{ \overline{\alpha}_{-1}=[0.1, 0.2],\; \overline{\alpha}_{+1}=[0.7, 0.9] \big\},   \label{eq:int_2} \\
        I_3 &= \big\{ \overline{\alpha}_{-1}=[0.0, 0.75],\; \overline{\alpha}_{+1}=[0.25, 1.0] \big\}. \label{eq:int_3} \\
        I_4 &= \big\{ \overline{\alpha}_{-1}=[0.0, 0.25) \cup [0.5, 0.75),\; \overline{\alpha}_{+1}=[0.25, 0.5) \cap [0.75, 1.0] \big\}. \label{eq:int_4}
    \end{align}
    In $I_1$, the classifier is trained to recognize if a given state $\rho_\alpha$ is taken from $\alpha < 0.5$ or $\alpha \geqslant 0.5$. In $I_2$, the classes are chosen to comprise subsets of $[0, 1]$ which do not overlap.
    The intervals $I_3$ are selected to test the performance of the classifier for intersecting regions.
    $I_4$ divides $[0,1]$ into four parts such that the first and the third parts belong to the class $y=-1$ and the second and the forth parts are marked by $y=+1$. The input state was set to $\rho^\mathrm{in} = \ketbra{+}{+}$, and the cost function \eqref{eq:kernel_cost} was maximized using the SLSQP method which supports bounds and constraints~\cite{2020SciPy-NMeth}. 
    
\begin{figure*}[h]
        \centering
        \includegraphics[width=.225\textwidth]{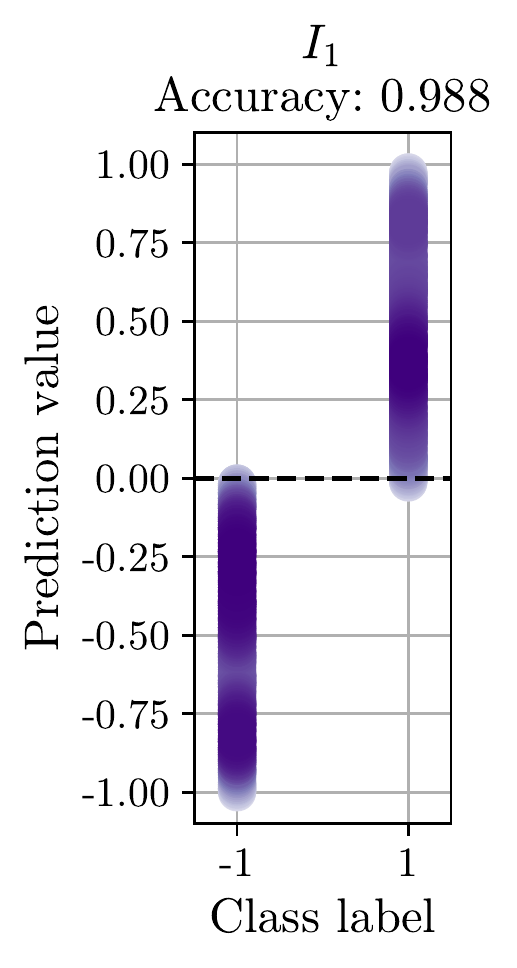}
        \includegraphics[width=.2145\textwidth]{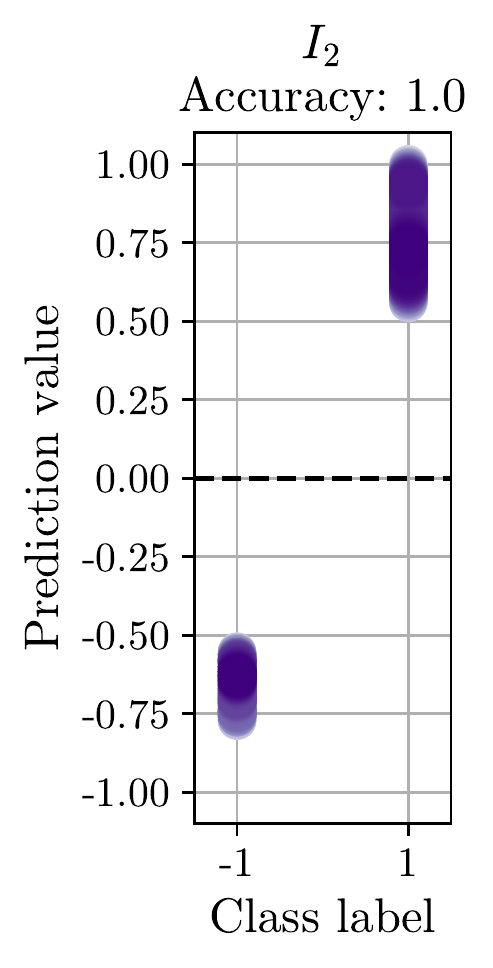}
        \includegraphics[width=.225\textwidth]{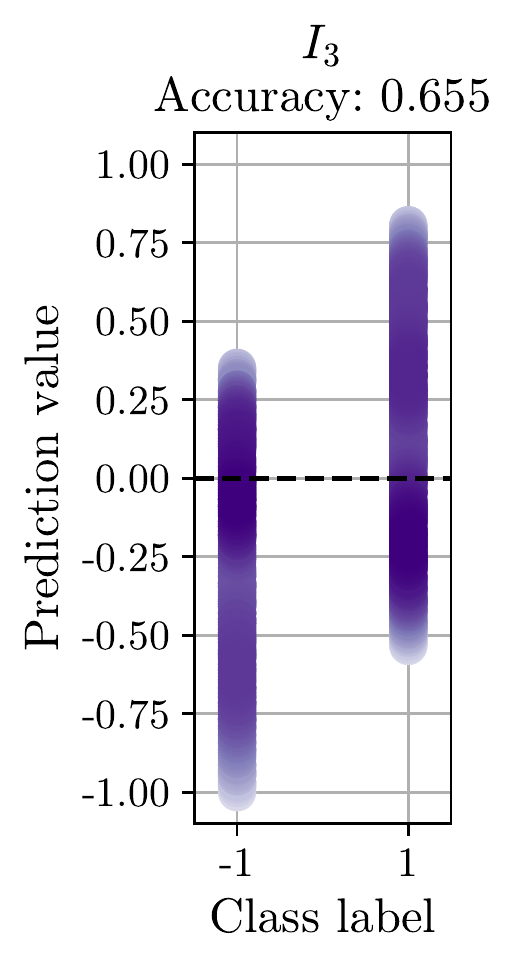}
        \includegraphics[width=.225\textwidth]{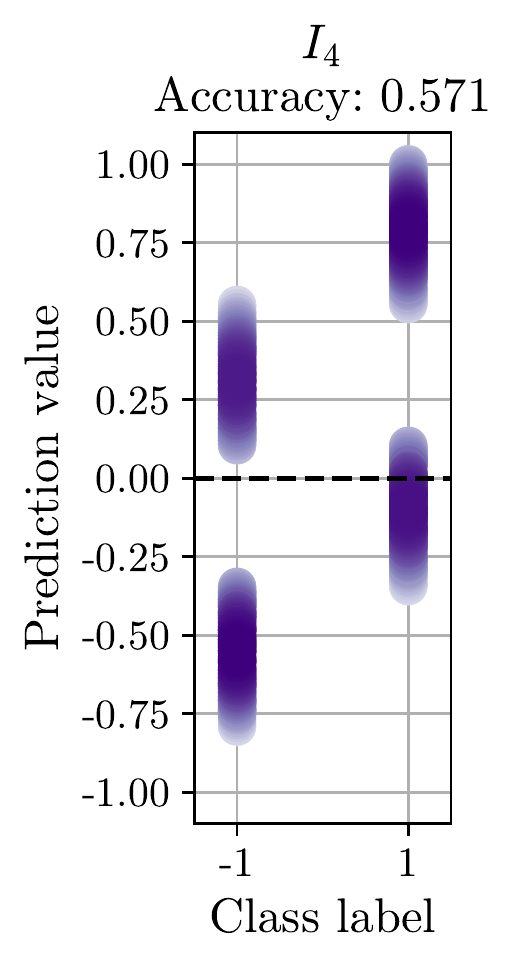}
        \caption{The accuracy of quantum classifiers trained to discriminate depolarizing channels corresponding to the intervals of the depolarization factor $I_1$, $I_2$, $I_3$ and $I_4$ and defined by Eqs.~\eqref{eq:int_1}, \eqref{eq:int_2}, \eqref{eq:int_3}, and \eqref{eq:int_4}, respectively. The input state is $\rho^\mathrm{in} = \ketbra{+}{+}$, and the sizes of the training and test sets are $N_{\text{train}} = 100$ and $N_{\text{test}} = 1000$. The vertical axis shows the normalized prediction value determined by~\eqref{eq:prediction_kernel}. The color intensity features the density of data points.}
        \label{fig:accuracy_kernels}
    \end{figure*}

    The results of our numerical simulations as presented in Fig.~\ref{fig:accuracy_kernels} reveal that the classifier trained to discriminate the channels from $I_1$ and $I_2$ provides excellent accuracy. Moreover, the higher accuracy is achieved in case the regions $\overline{\alpha}_{-1}, \overline{\alpha}_{+1}$ are separated. In contrast, we expect a priori low accuracy for $I_3$. For example, the states $\rho_{\alpha_j}$ from $[0.25, 0.75] = \overline{\alpha}_{-1} \cap \overline{\alpha}_{+1}$ could happen to have different labels $y$ for the same depolarization factor $\alpha$. 
    In this case, it turns out that the training performance is substantially dependent on the initial assignments of the parameters $\boldsymbol{\theta}$. 
    Remarkably, in case of $I_4$ the regions $\overline{\alpha}_{-1}$ and $\overline{\alpha}_{+1}$ do not intersect which translates to the fact that the states $\rho_j$ are expected to be classifiable, and yet the classifier fails.
    To remedy this issue, Alice could train the classifier on $n$ copies of the output states, $\rho_j^{\otimes n}$. This allows one to modify the kernel accordingly:
    \begin{equation}
        \label{eq:mod_ker}
        \mathcal{K}(\rho_i, \rho_j) = \Tr(\rho_i^{\otimes n} \rho_j^{\otimes n})= [\Tr(\rho_i \rho_j)]^n.
    \end{equation}
    The numerical results with this kernel are elaborated on in Appendix~\ref{app:mod_ker}.
    
\section{Discussion}

    To summarize, we discuss the approaches we used to solve the quantum channel discrimination problem. 
    First, we did put the task into the framework of variational quantum computing paradigm. 
    Namely, we stated the optimization problem of Eq.~\eqref{eq:p_suc} in terms of optimizing the parameters of an ansatz circuit, see \eqref{eq:psuc_par_var} and \eqref{eq:psuc_seq_var}.
    Potentially, this gives an opportunity for discriminating channels using noisy intermediate-scale quantum (NISQ) multi-qubit systems~\cite{preskill2018quantum}.
    In the context of variational quantum computing, we stressed out that the sequential strategy \eqref{eq:p_suc_seq} is superior to the parallel strategy \eqref{eq:p_suc_par} which is in line with the previous studies.
    The sequential strategy with $p=2$ channel applications allows one to perfectly discriminate the entanglement-breaking channels \eqref{eq:entbreak_ch}.
    In case of depolarizing channel~\eqref{eq:dep_channel}, the sequential strategy still performs better (see Fig.~\ref{fig:psucs-par-seq}), although the the total number of qubits to be used is the same for both methods.
    
    Being reformulated in terms of the variational quantum computing, the parallel strategy with $p$ channel applications requires a quantum computer of $p$ primary and $r$ ancillary qubits, so that the total amount of qubits is $Q_\mathrm{par}=p+r$. 
    On the other hand, in the sequential strategy, $Q_\mathrm{seq} = 1+r$ qubits have to be provided, revealing thus no dependence on $p$. 
    Despite the advantage in the number of qubits, with growing $p$ and $r$, the sequential strategy might be worse in terms of the number of optimization parameters $C = |\boldsymbol{\theta}|$. 
    Indeed, if every unitary $U(\boldsymbol{\theta}_k)$ is a hardware-efficient ansatz of $l$ layers, it necessitates $C_\mathrm{seq}=3l(1+r)(p+1) \sim O(pr)$ parameters to optimize over. 
    In contrary, in the parallel strategy, $C_\mathrm{par}=2 \cdot 3l(p+r) \sim O(p+r)$. 
    That is, by choosing a strategy one trades quantum resources $Q$ for classical resources $C$, and vice versa. 
    Recall that for $p=2$ our observation suggests that for the same total amount of qubits the sequential strategy outperforms the parallel strategy.
    
    In this work, we also addressed the quantum channel discrimination problem solved using a variational-circuit-based quantum classifier.
    It was mentioned that the best performance is achieved when the classifier is trained on the pairs of the original state $\rho$ and its copy $\Phi[\rho]$ which passed through a channel. 
    Inspired by the approach to quantum channel discrimination as realized with the use of parallel and sequential strategy, we attempted to train the classifier on the pairs of the state $\Phi[\rho]$ and the state $\ketbra{0}{0}^{\otimes r}$, so that the variational circuit is a $(r+1)$-layered hardware-efficient ansatz. 
    Furthermore, we performed the training on the $r$ copies of the state $\Phi[\rho]^{\otimes r}$. 
    However, none of these two training ways results in a good performance. Remarkably, training such a quantum classifier with a simpler and less expressive ansatz is advantageous. 
    In principle, the circuit of 7 parameters $U_1$ defined in~\eqref{eq:class_u1} is capable of preparing any pure two-qubit state, showing yet worse performance compared to the one trained with the ansatz $U_2$ in~\eqref{eq:class_u2} containing no entangling gates.
    We tried to add a $CX$ gate to this circuit, which increases its expressive power without introducing any optimization parameters. 
    Still the ansatz gives a lower performance, which suggests that this fact cannot be attributed to overparametrization.
    Despite the assumption that we are given a pair of $\rho$ and $\Phi[\rho]$ states, and we may, in principle, perform arbitrary number of measurements, this quantum machine learning approach is very powerful. 
    First, the original input states may be random and even mixed. 
    Second, one needs a circuit-based quantum computer of two qubits only and no entangling gates.
    
    Kernel-based methods for quantum channel discrimination were also studied in this work.
    We deliberately considered a more complex task with the channels being specified by the intervals $\overline{\alpha}_y$ of the depolarization factor $\alpha$ and not of its fixed values $\alpha_y$.
    The reason for this is that with fixed input states $\rho^\mathrm{in}$ and the two depolarizing channels with $\alpha_{\pm 1}$, it would be enough to have a training set of only $N_\mathrm{train}=2$ states, $\{\Phi(\alpha_{-1})[\rho^\mathrm{in}],  \Phi(\alpha_{+1})[\rho^\mathrm{in}]\}$. Special attention should be paid to the case of discriminating the depolarizing channels corresponding to the intervals $I_4$ in~\eqref{eq:int_4}, which divides the line $[0, 1]$ into four parts with assigning the class $y=-1$ to the 1st and 3d parts and the class $y=+1$ to the 2nd and 4th parts. In principle, these classes are expected to be separable, and yet our classifier fails to do that. In Appendix~\ref{app:mod_ker}, we show this issue can be relaxed by modifying the kernel as $\mathcal{K}(\rho_i, \rho_j) = \Tr (\rho_i^{\otimes n} \rho_j^{\otimes n}) = [\Tr(\rho_i \rho_j)]^n$ for $n\in \mathbb{N}$,
    which is similar to the classical kernel $\mathcal{K}(x_i, x_j) = (|\boldsymbol{x}_i \cdot \boldsymbol{x}_j |^2)^n$ for $\boldsymbol{x} \in \mathbb{R}^d$~\cite{schuld2021supervised}. 
    Moreover, this simple modification of the kernel allows one to use random and mixed input states $\rho^\mathrm{in}$ instead of the fixed $\rho^\mathrm{in} = \ketbra{+}{+}$.
    
    It should be stressed that machine learning tasks based on quantum kernel estimation are classical-quantum. 
    That is, one first maps classical data points $\boldsymbol{x} \in \mathbb{R}^d$ into pure quantum states $\ket{\boldsymbol{x}}$ of a Hilbert space, for which the density operators are $\rho = \ketbra{\boldsymbol{x}}{\boldsymbol{x}}$ and the kernel reduces to $\mathcal{K}(\boldsymbol{x}_i, \boldsymbol{x}_j) = |\langle\boldsymbol{x}_i|\boldsymbol{x}_j \rangle|^2$. 
    Such transformation is called a feature map, and in its simplest form it is specified by
    \begin{equation*}
        \boldsymbol{x} = \{x_i\}_{i=1}^d \quad\longrightarrow\quad \ket{\boldsymbol{x}} = \bigotimes_{i=1}^d \big(\cos(x_i/2) \ket{0} + \sin(x_i/2)\ket{1}\big).
    \end{equation*}
    In principle, one can suggest more efficient mapping scheme, meanwhile the necessity of encoding $\boldsymbol{x} \rightarrow \ket{\boldsymbol{x}}$ is considered as an important shortcoming of classical-quantum machine learning. However, in the task of quantum channel discrimination, the data points $\rho$ are quantum and do not need to be encoded, although these quantum states are in general mixed.
    
    Interestingly, the function $\mathcal{K}(\rho_i, \rho_j) = \Tr(\rho_i \rho_j)$ seems to play an important role in the other considered approaches of channel discrimination.
    That is, for the approach of variational computing embedding, we found that the less expressive the ansatz is (i.e. the fewer layers $l$ it has) the more it correlates with the trace of the product (see Appendix~\ref{app:trace_prod}). 
    In addition, for the variational quantum classifier, we observed that with a proper ansatz the classification is perfect for the depolarization factors $(\alpha_0, \alpha_1)$ such that $\alpha_0 \lesssim 0.75 \lesssim \alpha_1$. 
    For the output states of the depolarizing channel $\rho_\alpha = \Phi(\alpha)[\rho]$, the point $\alpha = 0.75$ is the extremum of the function $\mathcal{K}(\rho_\alpha) = \Tr\rho_\alpha^2$ with $\forall \rho \neq \mathbb{1}/2$ (or more generally, the minimum of $\mathcal{K}(\rho_\alpha, \rho_{\alpha+\epsilon}) = \Tr(\rho_\alpha \rho_{\alpha+\epsilon})$ is at $\alpha = 0.75 - \epsilon/2$), see Appendix~\ref{app:trace_prod} for details. By this we suggest that while solving the quantum channel discrimination problem, one must pay attention not only to the diamond-norm distance between the target channels, but also to the trace of the product of their output states.
    
    All the approaches considered in our study have their pros and cons, as well as different assumptions. 
    The first approach, variational circuit embedding assumes that we are given a number of channel applications $p$ and an arbitrary number of measurements in the training stage, but in the active stage only one measurement is allowed.
    On the other hand, the second approach, the variational quantum classifier assumes only a single channel application, but requires to be trained on the pairs of the output and original states $\rho_\alpha \otimes \rho$, and also needs many measurements for estimating expectation values. 
    But at the same time the states $\rho$ can be random and even mixed. The third approach which is based on quantum kernel estimation also requires many measurements for computing the kernel, but allows to discriminate parameter-dependent channels for different ranges of parameters that belong to different classes. As pointed out in Appendix~\ref{app:mod_ker}, this technique of quantum channel discrimination could be improved by training on $n$ copies of input states, which is equivalent to raising the kernel to the power of $n$. 
     
\section{Acknowledgments}
    
    This work was supported in the framework of the Roadmap for Quantum computing (Contract No. 868-1.3-15/15-2021 dated October 5, 2021 and Contract No. P2163/11148  dated December 3, 2021).
    DY acknowledges the support from the Russian Science Foundation Project No.~22-11-00074.
    The data and code that support the findings of this study are available from ASK upon reasonable request.
    
\onecolumngrid
\appendix
\newpage


\section{Dependence on the trace of the product} \label{app:trace_prod}

\begin{figure*}[h]
        \centering
        \includegraphics[width=.475\textwidth]{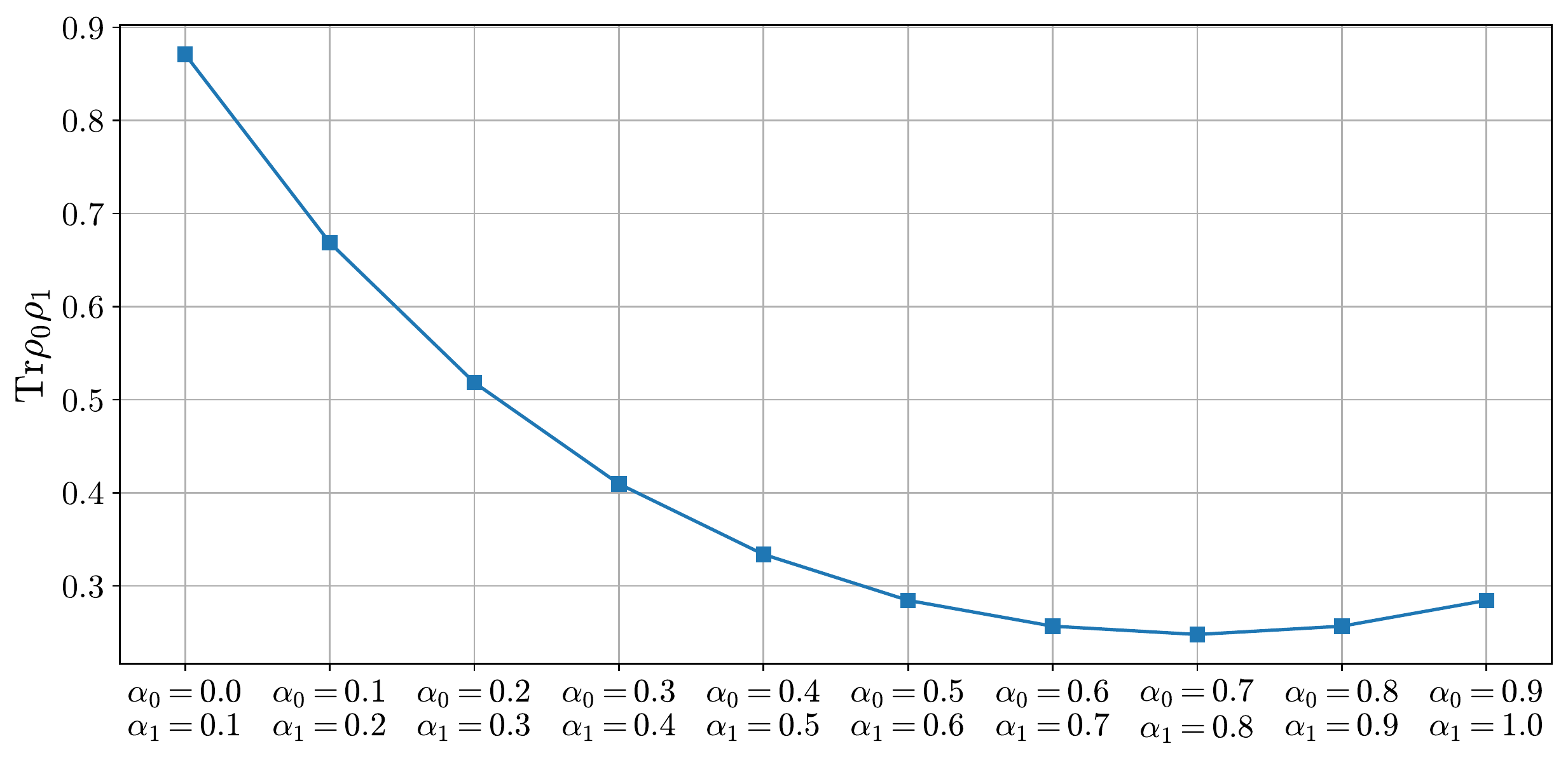}
        \includegraphics[width=.475\textwidth]{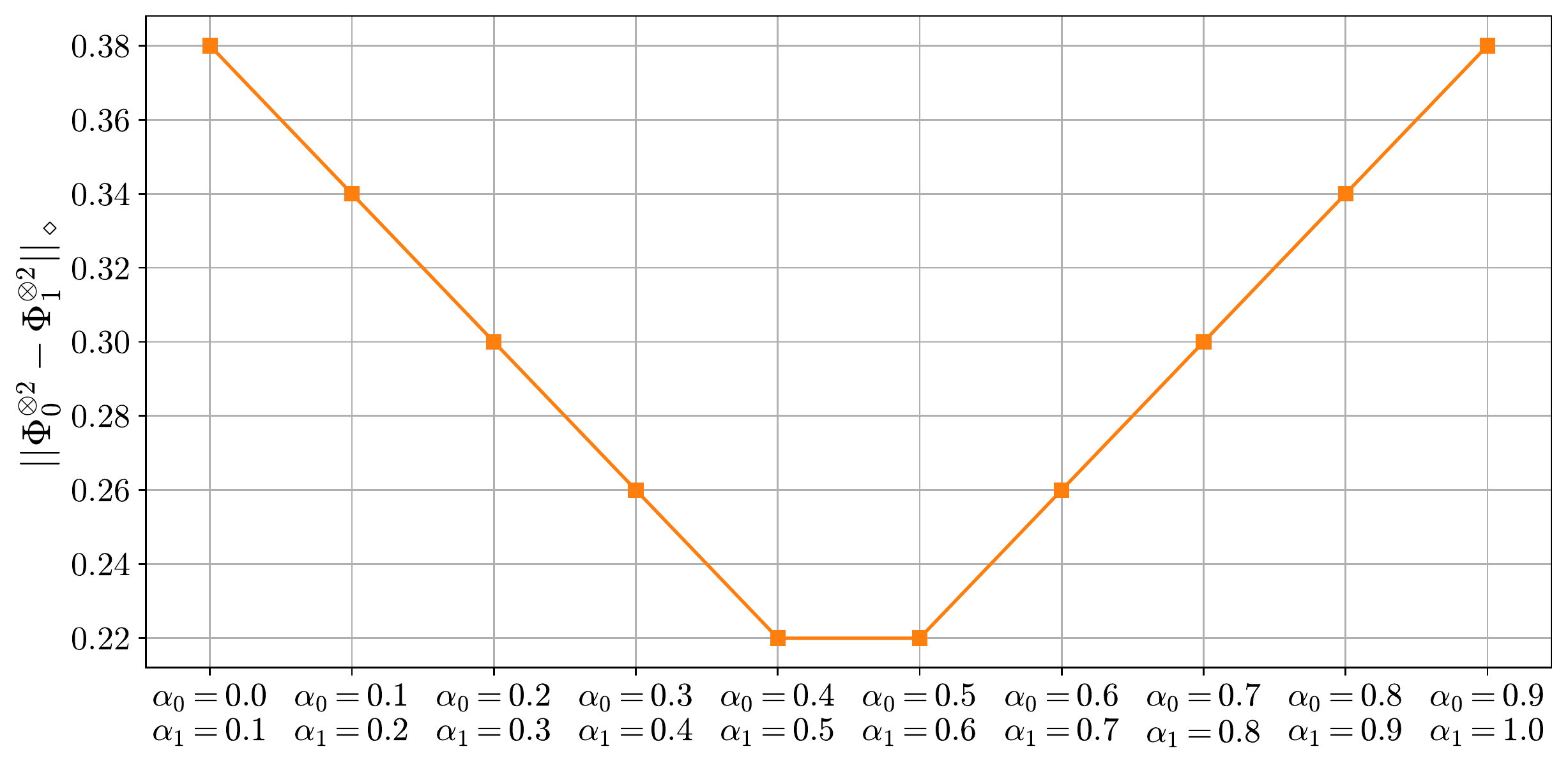}
        \caption{Trace of the product (left) and the diamond-distance (right) for the parallel discrimination strategy versus depolarization factors $(\alpha_0, \alpha_1)$ for $p=2$ channel applications. Herein, $\Phi_y \equiv \Phi(\alpha_y)$ is the depolarizing channel \eqref{eq:dep_channel}, and $\rho_y = \Phi(\alpha_y)[\rho]$.}
        \label{fig:tp-dd}
    \end{figure*}

\begin{figure*}[h]
        \centering
        \includegraphics[width=.475\textwidth]{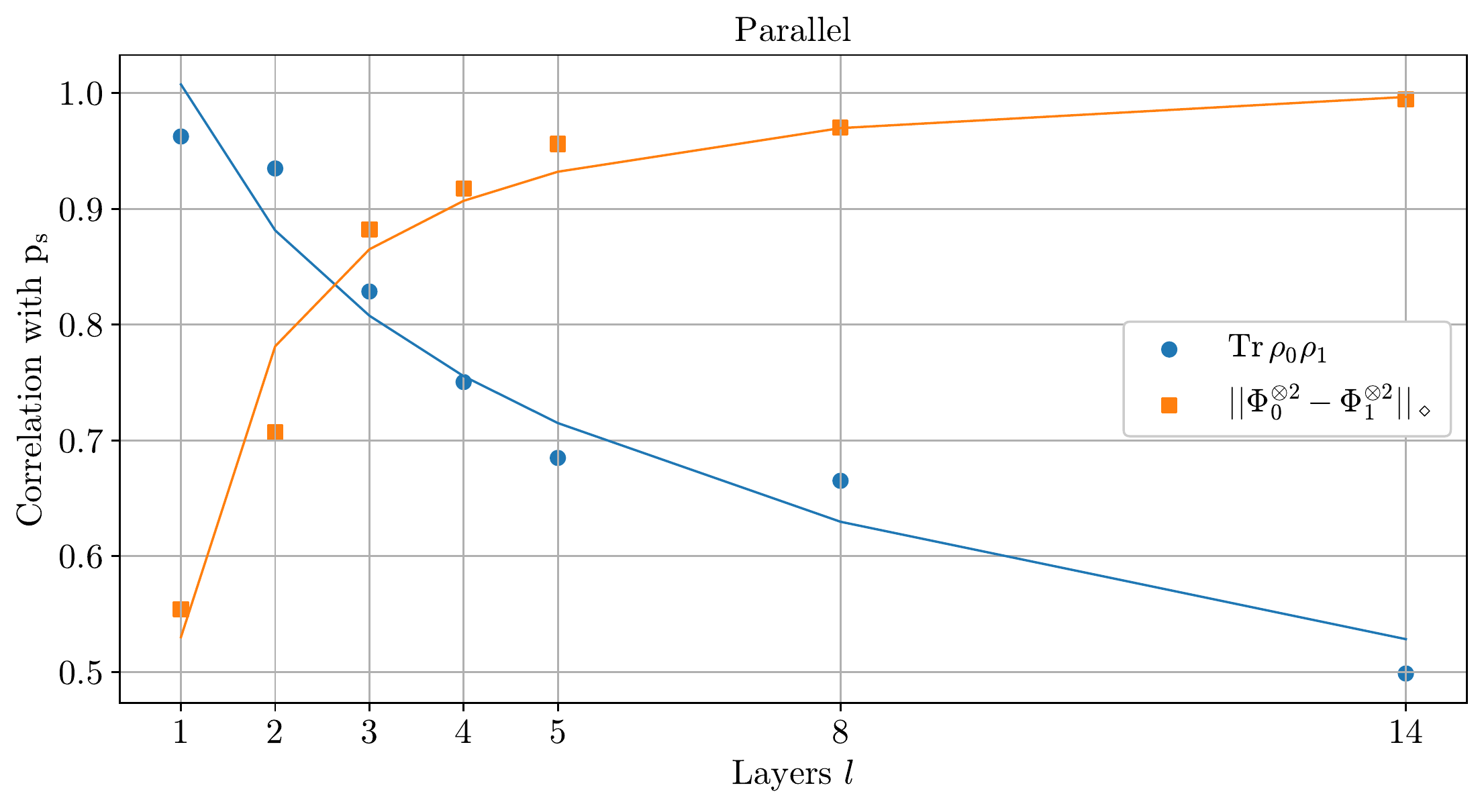}
        \includegraphics[width=.475\textwidth]{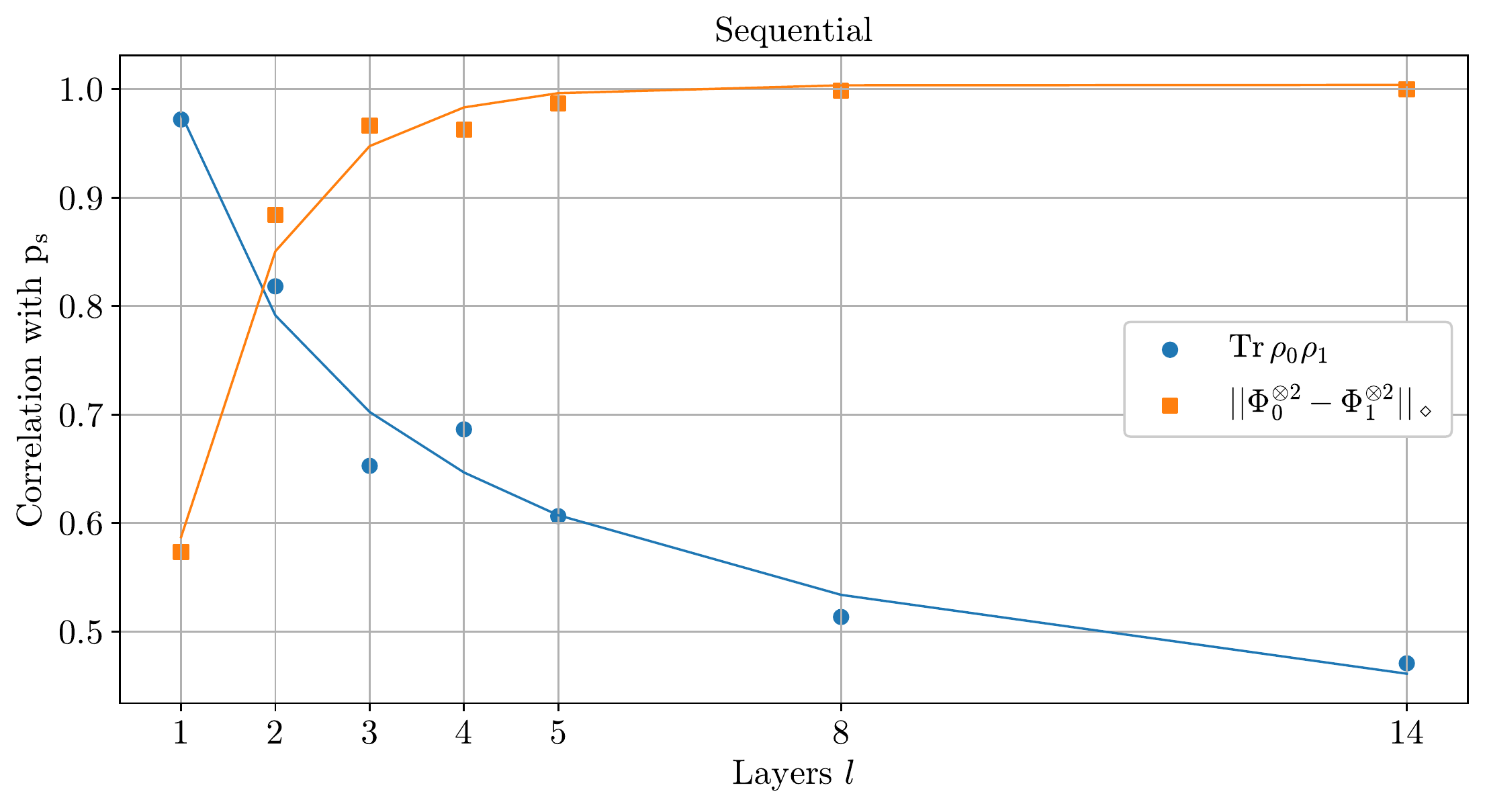}
        \caption{The Pearson correlation coefficients between the trace of the product, diamond distance, and the average successful discrimination probabilities. On the left are the results for the parallel strategy, and on the right are for the sequential strategy. Each data point is obtained by averaging out five independent runs. By solid lines, data points are fitted by the functions of $l$, the number of layers of the hardware-efficient ansatz. We used $f(l)=l^{-1/a}$ for fitting the trace of the product, and $g(l)=1-e^{-bl}$ for the diamond distance.
        }
        \label{fig:correlations}
    \end{figure*}

    While solving the quantum channel discrimination problem in the varialtional quantum computing framework (see \eqref{eq:psuc_par_var} and \eqref{eq:psuc_seq_var} in the main text), one may expect that the performance significantly depends on the number of ansatz layers $l$ and on the diamond-norm distance between the channels which determines $\mathrm{p_\diamond^{par}}$ as given in~\eqref{eq:p_suc_diamond_par}.
    Considering the depolarizing channels $\Phi_0$ and $\Phi_1$ in the form of \eqref{eq:dep_channel} with the depolarization factors $\alpha_0$ and $\alpha_1$, it can be seen that the diamond-distance $||\Phi^{\otimes p}_0 - \Phi_1^{\otimes p}||_\diamond$ for $p=2$ channel applications is symmetric (see Fig.~\ref{fig:tp-dd}). At the same time, in our numerical simulations, we observe that it is harder to achieve theoretical success probability $\mathrm{p_\diamond^{par}}$ for the pairs of factors $(\alpha_0, \alpha_1)$ which are on the right side to $\alpha=0.5$ (see Fig.~\ref{fig:psucs-par-seq}).
    Interestingly, this seems to correlate with the trace of the product of density operators $\mathrm{Tr}(\rho_0 \rho_1)$, where $\rho_y = \Phi(\alpha_y)[\rho]$, $\forall \rho \neq \mathbb{1}/2$ (see Fig.~\ref{fig:tp-dd}). 
    Moreover, it appears that the more ansatz layers $l$ one uses to maximize the success probability $\mathrm{p_s}$, the less the convergence properties depend on the trace of the product. 
    In Fig.~\ref{fig:correlations}, this can be observed upon close inspection of the Pearson's coefficients.

\section{Modified kernel} \label{app:mod_ker}

    In machine learning, the kernel is a complex- or real-valued function $\mathcal{K}(\rho_i, \rho_j)$ that produces a positive semi-definite matrix $\mathcal{K}_{ij} = \mathcal{K}(\rho_i, \rho_j)$. 
    Among various kernels considered in the domain of classical machine learning, the simplest one is $\mathcal{K}(\boldsymbol{x}_i, \boldsymbol{x}_j) = (|\boldsymbol{x}_i \cdot \boldsymbol{x}_j|^2)^n$ for some $n \in \mathbb{N}$ and classical data $\boldsymbol{x} \in \mathbb{R}^d$. 
    When the data is quantum, i.e. $\rho$ is a density operator on $\mathbb{C}^d$, a similar kernel is $\mathcal{K}(\rho_i, \rho_j) = \Tr (\rho_i^{\otimes n} \rho_j^{\otimes n}) = [\Tr (\rho_i \rho_j) ]^n$ given in \eqref{eq:mod_ker} in the main text.
    
\begin{figure*}[h]
        \centering
        \includegraphics[width=.225\textwidth]{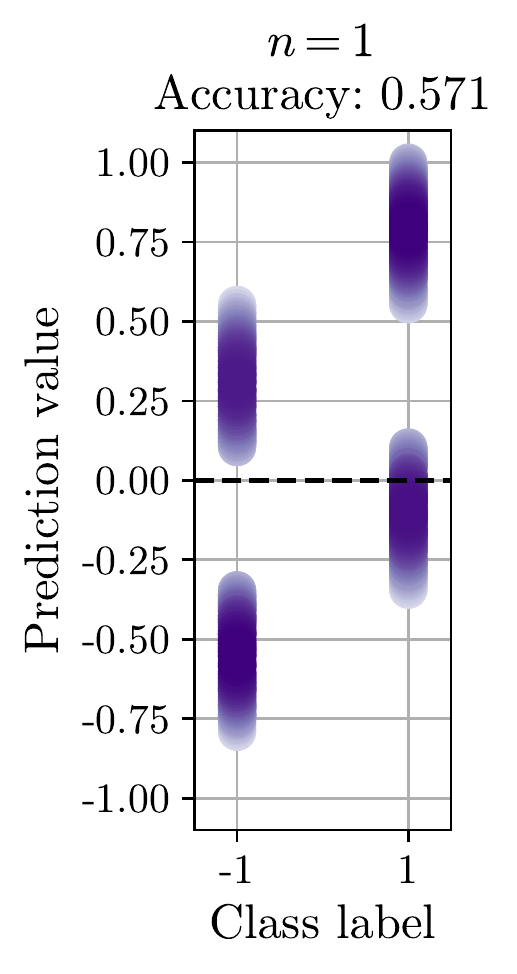}
        \includegraphics[width=.2195\textwidth]{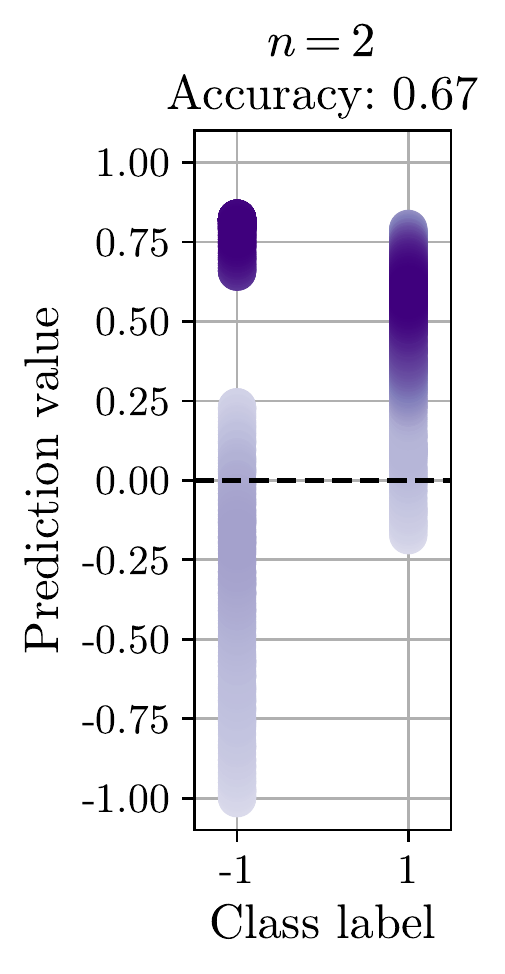}
        \includegraphics[width=.225\textwidth]{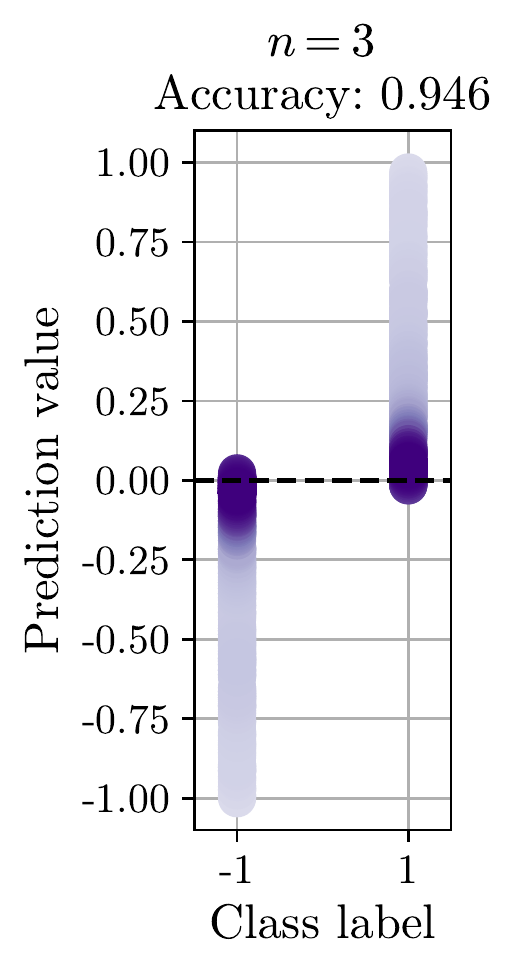}
        \includegraphics[width=.225\textwidth]{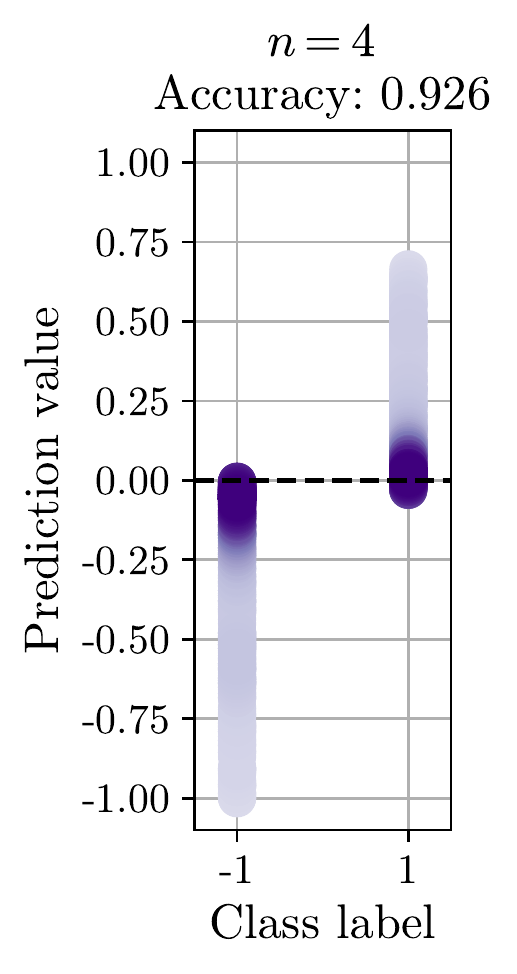}
        \caption{The accuracy of the kernel-based classifier obtained on the test set for the intervals $I_4$ defined in \eqref{eq:int_4}.
        From left to right given are the accuracy for the classifiers based on the kernel \eqref{eq:mod_ker} for different $n \in \{1, 2, 3, 4\}$.
        The input state is $\rho^\mathrm{in} = \ketbra{+}{+}$, and the sizes of the training and test sets are $N_{\text{train}} = 100$ and $N_{\text{test}} = 1000$.
        The vertical axis shows the normalized prediction value defined in \eqref{eq:prediction_kernel}.
        The color intensity features the density of data points.}
        \label{fig:accuracy_kernels_mod}
    \end{figure*}

    In our numerical simulations, we can see that the performance of the kernel-based classifier may depend on $n$, the number of copies of the channel output states used to train the classifier. That is, in Fig.~\ref{fig:accuracy_kernels} one can notice that for the intervals $$I_4 = \big\{ \overline{\alpha}_{-1}=[0.0, 0.25) \cup [0.5, 0.75),\; \overline{\alpha}_{+1}=[0.25, 0.5) \cap [0.75, 1.0] \big\}$$ mentioned in \eqref{eq:int_4}, the classifier trained on the single-copy states ($n=1$) fails to separate the classes. In Fig.~\ref{fig:accuracy_kernels_mod}, we show the results of classification for $I_4$ obtained with different numbers of state copies $n$. Apparently, for achieving the best classification accuracy in this case one should use $n=3$ copies of the channel output states for training. In this case, one modifies the kernel $\mathcal{K}(\rho_i, \rho_j) = \Tr (\rho_i \rho_j)$ such that it is just raised to the power of $n=3$.

\begin{figure*}[h]
        \centering
        \includegraphics[width=.225\textwidth]{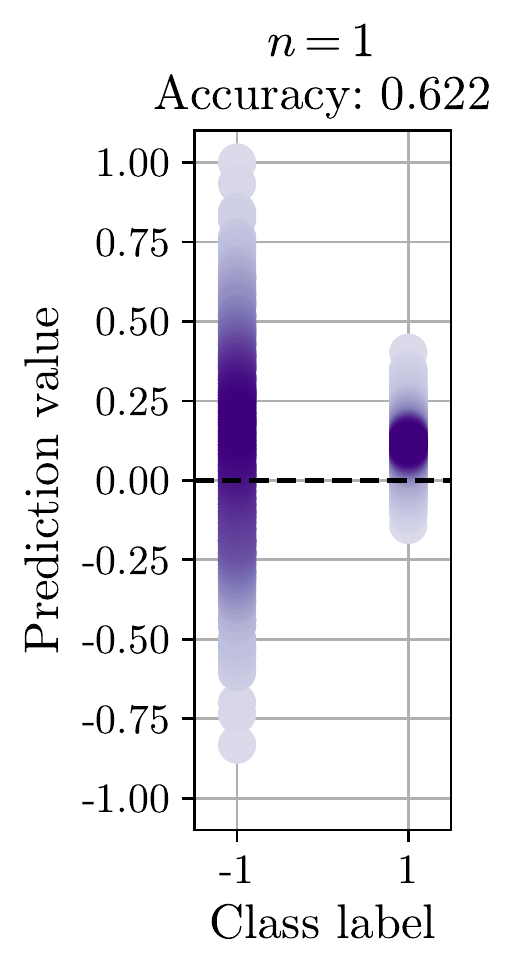}
        \includegraphics[width=.225\textwidth]{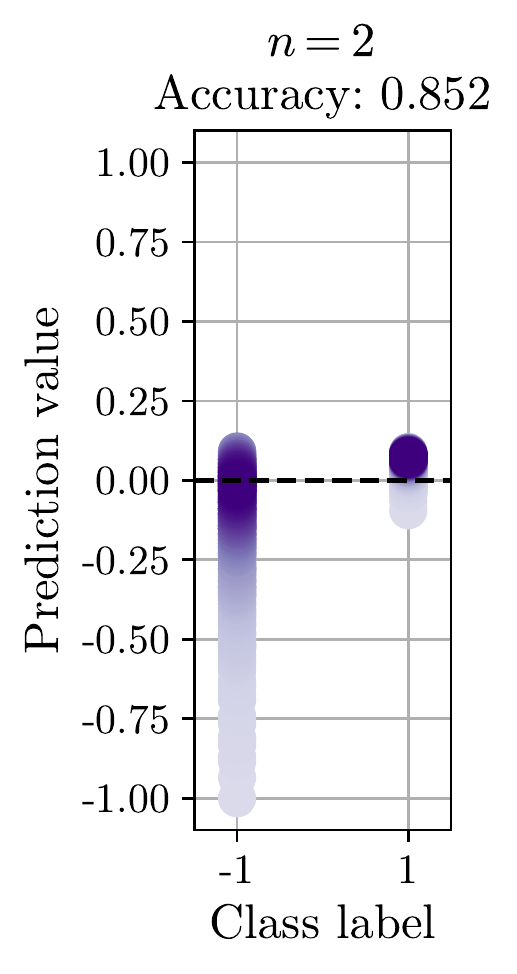}
        \includegraphics[width=.225\textwidth]{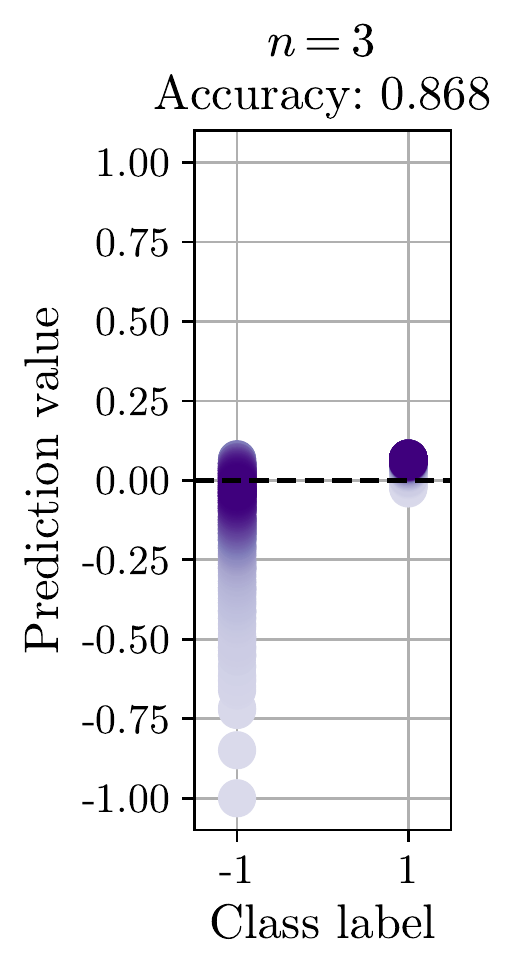}
        \includegraphics[width=.2195\textwidth]{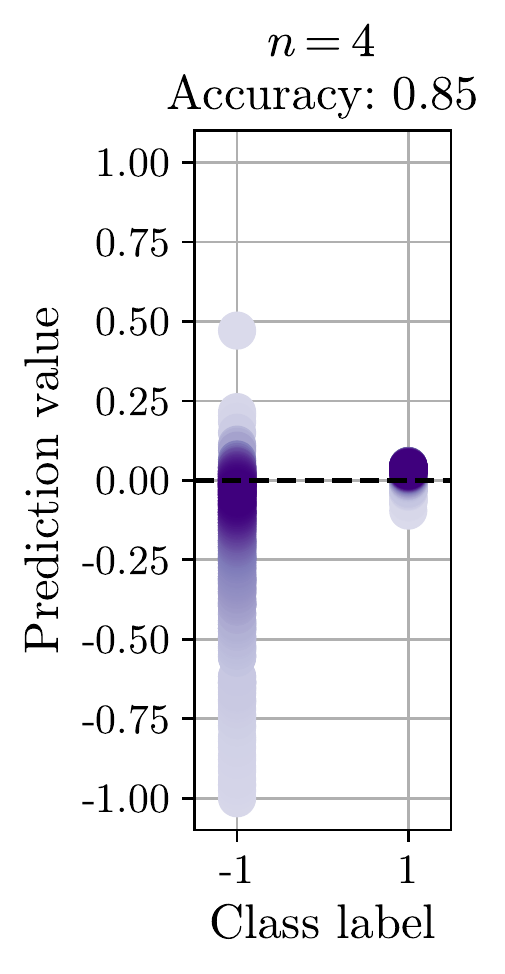}
        \caption{The accuracy of the kernel-based classifier obtained on the test set for the intervals $I_1$ defined in~\eqref{eq:int_1}.
        From left to right given are the accuracy for the classifiers based on the kernel~\eqref{eq:mod_ker} for different $n \in \{1, 2, 3, 4\}$. The input states $\rho^\mathrm{in}$ are random and mixed, and the sizes of the training and test sets are $N_{\text{train}} = 100$ and $N_{\text{test}} = 1000$. The vertical axis shows the normalized prediction value defined in~\eqref{eq:prediction_kernel}. The color intensity features the density of data points.}
        \label{fig:accuracy_kernels_mod_rand}
    \end{figure*}

    As mentioned in the main text, we also discovered that this modification of the kernel makes the classifier more powerful in terms of the allowed input states. That is, instead of $\rho^\mathrm{in} = \ketbra{+}{+}$, the input state can be random and mixed, as for the variational quantum classifier we tested in our study. In Fig.~\ref{fig:accuracy_kernels_mod_rand}, we show classification accuracy for the intervals $$I_1 = \big\{ \overline{\alpha}_{-1}=[0.0, 0.5),\; \overline{\alpha}_{+1}=[0.5, 1.0]\big\}$$ and random mixed input states $(\rho^\mathrm{in})^{\otimes n}$ with $n \in \{1, 2, 3, 4\}$. As can be seen, with $n=1$ the classifier fails to predict the labels when the input states are random, compared to the case when they are fixed (recall that the intervals $I_1$ are discussed in \eqref{eq:int_1} and tested for the classifier with fixed input $\rho^\mathrm{in} = \ketbra{+}{+}$, see Fig.~\ref{fig:accuracy_kernels}).
    
\newpage
    
\bibliography{bibliography}

\providecommand{\noopsort}[1]{}\providecommand{\singleletter}[1]{#1}
\begin{thebibliography}{10}

\bibitem{kitaev1997quantum}
A.~Yu. Kitaev.
\newblock Quantum computations: algorithms and error correction.
\newblock {\em Russian Mathematical Surveys}, 52(6):1191, 1997.

\bibitem{wang2006unambiguous}
G.~Wang and M.~Ying.
\newblock Unambiguous discrimination among quantum operations.
\newblock {\em Physical Review A}, 73(4):042301, 2006.

\bibitem{duan2009perfect}
R.~Duan, Y.~Feng, and M.~Ying.
\newblock Perfect distinguishability of quantum operations.
\newblock {\em Physical Review Letters}, 103(21):210501, 2009.

\bibitem{Holevo_2012}
A.~Holevo.
\newblock {\em Quantum Systems, Channels, Information: A Mathematical
  Introduction}.
\newblock de Gruyter \& Co, Berlin, Boston, 2012.

\bibitem{pirandola2018advances}
S.~Pirandola, B.~R. Bardhan, T.~Gehring, C.~Weedbrook, and S.~Lloyd.
\newblock Advances in photonic quantum sensing.
\newblock {\em Nature Photonics}, 12(12):724--733, 2018.

\bibitem{lloyd2008enhanced}
S.~Lloyd.
\newblock Enhanced sensitivity of photodetection via quantum illumination.
\newblock {\em Science}, 321(5895):1463--1465, 2008.

\bibitem{barzanjeh2015microwave}
S.~Barzanjeh, S.~Guha, C.~Weedbrook, D.~Vitali, J.~H. Shapiro, and
  S.~Pirandola.
\newblock Microwave quantum illumination.
\newblock {\em Physical Review Letters}, 114(8):080503, 2015.

\bibitem{ortolano2021experimental}
Giuseppe Ortolano, Elena Losero, Stefano Pirandola, Marco Genovese, and Ivano
  Ruo-Berchera.
\newblock Experimental quantum reading with photon counting.
\newblock {\em Science Advances}, 7(4):eabc7796, 2021.

\bibitem{Brandt_1999}
H.~E. Brandt.
\newblock Positive operator valued measure in quantum information processing.
\newblock {\em American Journal of Physics}, 67(5):434--439, 1999.

\bibitem{James_2001}
D.~F.~V. James, P.~G. Kwiat, W.~J. Munro, and A.~G. White.
\newblock Measurement of qubits.
\newblock {\em Physical Review A}, 64:052312, Oct 2001.

\bibitem{Palmieri_2020}
A.~M. Palmieri, E.~Kovlakov, F.~Bianchi, D.~Yudin, S.~Straupe, J.~D. Biamonte,
  and S.~Kulik.
\newblock Experimental neural network enhanced quantum tomography.
\newblock {\em npj Quantum Information}, 6:20, 2020.

\bibitem{harrow2010adaptive}
A.~W Harrow, A.~Hassidim, D.~W Leung, and J.~Watrous.
\newblock Adaptive versus nonadaptive strategies for quantum channel
  discrimination.
\newblock {\em Physical Review A}, 81(3):032339, 2010.

\bibitem{wilde2020coherent}
M.~M Wilde.
\newblock Coherent quantum channel discrimination.
\newblock In {\em 2020 IEEE International Symposium on Information Theory
  (ISIT)}, pages 1915--1920. IEEE, 2020.

\bibitem{farooq2018quantum}
J.~ur~Rehman, A.~Farooq, Y.~Jeong, and H.~Shin.
\newblock Quantum channel discrimination without entanglement.
\newblock {\em Quantum Information Processing}, 17(10):1--16, 2018.

\bibitem{bavaresco2021strict}
J.~Bavaresco, M.~Murao, and M.~T. Quintino.
\newblock Strict hierarchy between parallel, sequential, and
  indefinite-causal-order strategies for channel discrimination.
\newblock {\em Physical Review Letters}, 127(20):200504, 2021.

\bibitem{cooney2016strong}
Tom Cooney, Mil{\'a}n Mosonyi, and Mark~M Wilde.
\newblock Strong converse exponents for a quantum channel discrimination
  problem and quantum-feedback-assisted communication.
\newblock {\em Communications in Mathematical Physics}, 344(3):797--829, 2016.

\bibitem{katariya2021geometric}
V.~Katariya and M.~M Wilde.
\newblock Geometric distinguishability measures limit quantum channel
  estimation and discrimination.
\newblock {\em Quantum Information Processing}, 20(2):1--170, 2021.

\bibitem{Pirandola_2019}
S.~Pirandola, R.~Laurenza, C.~Lupo, and J.~L. Pereira.
\newblock Fundamental limits to quantum channel discrimination.
\newblock {\em npj Quantum Information}, 5(1), 2019.

\bibitem{pereira2021bounds}
J.~L Pereira and S.~Pirandola.
\newblock Bounds on amplitude-damping-channel discrimination.
\newblock {\em Physical Review A}, 103(2):022610, 2021.

\bibitem{zhuang2020ultimate}
Q.~Zhuang and S.~Pirandola.
\newblock Ultimate limits for multiple quantum channel discrimination.
\newblock {\em Physical Review Letters}, 125(8):080505, 2020.

\bibitem{wilde2020amortized}
M.~M. Wilde, M.~Berta, C.~Hirche, and E.~Kaur.
\newblock Amortized channel divergence for asymptotic quantum channel
  discrimination.
\newblock {\em Letters in Mathematical Physics}, 110(8):2277--2336, 2020.

\bibitem{katariya2021evaluating}
V.~Katariya and M.~M Wilde.
\newblock Evaluating the advantage of adaptive strategies for quantum channel
  distinguishability.
\newblock {\em Physical Review A}, 104(5):052406, 2021.

\bibitem{salek2020adaptive}
F.~Salek, M.~Hayashi, and A.~Winter.
\newblock When are adaptive strategies in asymptotic quantum channel
  discrimination useful?
\newblock {\em arXiv preprint arXiv:2011.06569}, 2020.

\bibitem{Harney_2021}
C.~Harney and S.~Pirandola.
\newblock Idler-free multi-channel discrimination via multipartite probe
  states.
\newblock {\em npj Quantum Information}, 7(1):153, 2021.

\bibitem{mcclean2016theory}
J.~R. McClean, J.~Romero, R.~Babbush, and A.~Aspuru-Guzik.
\newblock The theory of variational hybrid quantum-classical algorithms.
\newblock {\em New Journal of Physics}, 18(2):023023, 2016.

\bibitem{li2017hybrid}
J.~Li, X.~Yang, X.~Peng, and C.-P. Sun.
\newblock Hybrid quantum-classical approach to quantum optimal control.
\newblock {\em Physical Review Letters}, 118(15):150503, 2017.

\bibitem{santagati2018witnessing}
R.~Santagati et~al.
\newblock Witnessing eigenstates for quantum simulation of hamiltonian spectra.
\newblock {\em Science Advances}, 4(1):eaap9646, 2018.

\bibitem{peruzzo2014variational}
A.~Peruzzo et~al.
\newblock A variational eigenvalue solver on a photonic quantum processor.
\newblock {\em Nature Communications}, 5:4213, 2014.

\bibitem{biamonte2021universal}
Jacob Biamonte.
\newblock Universal variational quantum computation.
\newblock {\em Physical Review A}, 103(3):L030401, 2021.

\bibitem{cerezo2021variational}
M.~Cerezo et~al.
\newblock Variational quantum algorithms.
\newblock {\em Nature Reviews Physics}, 3(9):625--644, 2021.

\bibitem{larose2019variational}
R.~LaRose, A.~Tikku, {\'E}.~O’Neel-Judy, L.~Cincio, and P.~J Coles.
\newblock Variational quantum state diagonalization.
\newblock {\em npj Quantum Information}, 5(1):1--10, 2019.

\bibitem{lubasch2020variational}
M.~Lubasch, J.~Joo, P.~Moinier, M.~Kiffner, and D.~Jaksch.
\newblock Variational quantum algorithms for nonlinear problems.
\newblock {\em Physical Review A}, 101(1):010301, 2020.

\bibitem{biamonte2017quantum}
J.~Biamonte, P.~Wittek, N.~Pancotti, P.~Rebentrost, N.~Wiebe, and S.~Lloyd.
\newblock Quantum machine learning.
\newblock {\em Nature}, 549(7671):195--202, 2017.

\bibitem{huggins2019towards}
W.~Huggins, P.~Patil, B.~Mitchell, K.~B. Whaley, and E.~M. Stoudenmire.
\newblock Towards quantum machine learning with tensor networks.
\newblock {\em Quantum Science and Technology}, 4(2):024001, 2019.

\bibitem{schuld2015introduction}
M.~Schuld, I.~Sinayskiy, and F.~Petruccione.
\newblock An introduction to quantum machine learning.
\newblock {\em Contemporary Physics}, 56(2):172--185, 2015.

\bibitem{schuld2021machine}
M.~Schuld and F.~Petruccione.
\newblock {\em Machine Learning with Quantum Computers}.
\newblock Springer, 2021.

\bibitem{schuld2019quantum}
M.~Schuld and N.~Killoran.
\newblock Quantum machine learning in feature hilbert spaces.
\newblock {\em Physical Review Letters}, 122(4):040504, 2019.

\bibitem{mitarai2018quantum}
K.~Mitarai, M.~Negoro, M.~Kitagawa, and K.~Fujii.
\newblock Quantum circuit learning.
\newblock {\em Physical Review A}, 98(3):032309, 2018.

\bibitem{schuld2020circuit}
M.~Schuld, A.~Bocharov, K.~M Svore, and N.~Wiebe.
\newblock Circuit-centric quantum classifiers.
\newblock {\em Physical Review A}, 101(3):032308, 2020.

\bibitem{chen2020variational}
S.~Y.-C. Chen, C.-H.~H. Yang, J.~Qi, P.-Y. Chen, X.~Ma, and H.-S. Goan.
\newblock Variational quantum circuits for deep reinforcement learning.
\newblock {\em IEEE Access}, 8:141007--141024, 2020.

\bibitem{lloyd2020quantum}
S.~Lloyd, M.~Schuld, A.~Ijaz, J.~Izaac, and N.~Killoran.
\newblock Quantum embeddings for machine learning.
\newblock {\em arXiv preprint arXiv:2001.03622}, 2020.

\bibitem{uvarov2020machine}
A.~V. Uvarov, A.~S. Kardashin, and J.~D Biamonte.
\newblock Machine learning phase transitions with a quantum processor.
\newblock {\em Physical Review A}, 102(1):012415, 2020.

\bibitem{Uvarov_2020}
A.~Uvarov, J.~D. Biamonte, and D.~Yudin.
\newblock Variational quantum eigensolver for frustrated quantum systems.
\newblock {\em Physical Review B}, 102:075104, Aug 2020.

\bibitem{Kardashin_2020}
A.~Kardashin, A.~Uvarov, D.~Yudin, and J.~Biamonte.
\newblock Certified variational quantum algorithms for eigenstate preparation.
\newblock {\em Physical Review A}, 102:052610, Nov 2020.

\bibitem{Kardashin_2021}
A.~Kardashin, A.~Pervishko, J.~Biamonte, and D.~Yudin.
\newblock Numerical hardware-efficient variational quantum simulation of a
  soliton solution.
\newblock {\em Physical Review A}, 104:L020402, Aug 2021.

\bibitem{lazzarin2022multi}
M.~Lazzarin, D.~E. Galli, and E.~Prati.
\newblock Multi-class quantum classifiers with tensor network circuits for
  quantum phase recognition.
\newblock {\em Physics Letters A}, page 128056, 2022.

\bibitem{patterson2021quantum}
A.~Patterson, H.~Chen, L.~Wossnig, S.~Severini, D.~Browne, and I.~Rungger.
\newblock Quantum state discrimination using noisy quantum neural networks.
\newblock {\em Physical Review Research}, 3(1):013063, 2021.

\bibitem{chen2021universal}
H.~Chen, L.~Wossnig, S.~Severini, H.~Neven, and M.~Mohseni.
\newblock Universal discriminative quantum neural networks.
\newblock {\em Quantum Machine Intelligence}, 3(1):1--11, 2021.

\bibitem{mengoni2019kernel}
R.~Mengoni and A.~Di~Pierro.
\newblock Kernel methods in quantum machine learning.
\newblock {\em Quantum Machine Intelligence}, 1(3):65--71, 2019.

\bibitem{yu2019reconstruction}
S.~Yu et~al.
\newblock Reconstruction of a photonic qubit state with reinforcement learning.
\newblock {\em Advanced Quantum Technologies}, 2(7-8):1800074, 2019.

\bibitem{horodecki2003entanglement}
M.~Horodecki, P.~W. Shor, and M.~B. Ruskai.
\newblock Entanglement breaking channels.
\newblock {\em Reviews in Mathematical Physics}, 15(06):629--641, 2003.

\bibitem{ruskai2003qubit}
M.~B. Ruskai.
\newblock Qubit entanglement breaking channels.
\newblock {\em Reviews in Mathematical Physics}, 15(06):643--662, 2003.

\bibitem{shaham2015entanglement}
A.~Shaham, A.~Halevy, L.~Dovrat, E.~Megidish, and H.~S. Eisenberg.
\newblock Entanglement dynamics in the presence of controlled unital noise.
\newblock {\em Scientific Reports}, 5(1):1--8, 2015.

\bibitem{shaham2012realizing}
A.~Shaham and H.~S. Eisenberg.
\newblock Realizing a variable isotropic depolarizer.
\newblock {\em Optics Letters}, 37(13):2643--2645, 2012.

\bibitem{benenti2010computing}
G.~Benenti and G.~Strini.
\newblock Computing the distance between quantum channels: usefulness of the
  fano representation.
\newblock {\em Journal of Physics B: Atomic, Molecular and Optical Physics},
  43(21):215508, 2010.

\bibitem{puzzuoli2017ancilla}
D.~Puzzuoli and J.~Watrous.
\newblock Ancilla dimension in quantum channel discrimination.
\newblock In {\em Annales Henri Poincar{\'e}}, volume~18, pages 1153--1184.
  Springer, 2017.

\bibitem{matthews2010entanglement}
W.~Matthews, M.~Piani, and J.~Watrous.
\newblock Entanglement in channel discrimination with restricted measurements.
\newblock {\em Physical Review A}, 82(3):032302, 2010.

\bibitem{chiribella2008quantum}
G.~Chiribella, G.~M. D’Ariano, and P.~Perinotti.
\newblock Quantum circuit architecture.
\newblock {\em Physical Review Letters}, 101(6):060401, 2008.

\bibitem{qiu2019solving}
P.-H. Qiu, X.-G. Chen, and Y.-W. Shi.
\newblock Solving quantum channel discrimination problem with quantum networks
  and quantum neural networks.
\newblock {\em IEEE Access}, 7:50214--50222, 2019.

\bibitem{biamonte2019lectures}
J.~Biamonte.
\newblock Lectures on quantum tensor networks.
\newblock {\em arXiv preprint arXiv:1912.10049}, 2019.

\bibitem{nielsen2002quantum}
M.~A. Nielsen and I.~Chuang.
\newblock {\em Quantum Computation and Quantum Information}.
\newblock Cambridge University Press, 2010.

\bibitem{kandala_hardware-efficient_2017}
A.~Kandala, A.~Mezzacapo, K.~Temme, M.~Takita, M.~Brink, J.~M. Chow, and J.~M.
  Gambetta.
\newblock Hardware-efficient variational quantum eigensolver for small
  molecules and quantum magnets.
\newblock {\em Nature}, 549:242, Sep 2017.

\bibitem{watrous2018theory}
J.~Watrous.
\newblock {\em The Theory of Quantum Information}.
\newblock Cambridge University Press, 2018.

\bibitem{agrawal2018rewriting}
A.~Agrawal, R.~Verschueren, S.~Diamond, and S.~Boyd.
\newblock A rewriting system for convex optimization problems.
\newblock {\em Journal of Control and Decision}, 5(1):42--60, 2018.

\bibitem{diamond2016cvxpy}
S.~Diamond and S.~Boyd.
\newblock {CVXPY}: {A} {P}ython-embedded modeling language for convex
  optimization.
\newblock {\em Journal of Machine Learning Research}, 17(83):1--5, 2016.

\bibitem{byrd1995limited}
R.~H. Byrd, P.~Lu, J.~Nocedal, and C.~Zhu.
\newblock A limited memory algorithm for bound constrained optimization.
\newblock {\em SIAM Journal on scientific computing}, 16(5):1190--1208, 1995.

\bibitem{Shirinyan_2019}
A.~A. Shirinyan, V.~K. Kozin, J.~Hellsvik, M.~Pereiro, O.~Eriksson, and
  D.~Yudin.
\newblock Self-organizing maps as a method for detecting phase transitions and
  phase identification.
\newblock {\em Physical Review B}, 99:041108, Jan 2019.

\bibitem{Berezutskii_2020}
A.~Berezutskii, M.~Beketov, D.~Yudin, Z.~Zimbor{\'{a}}s, and J.~D. Biamonte.
\newblock Probing criticality in quantum spin chains with neural networks.
\newblock {\em Journal of Physics: Complexity}, 1(3):03LT01, aug 2020.

\bibitem{scholkopf2002learning}
B.~Sch{\"o}lkopf and A.~J. Smola.
\newblock {\em Learning with Kernels: Support Vector Machines, Regularization,
  Optimization, and Beyond}.
\newblock The MIT Press, 2018.

\bibitem{steinwart2008support}
I.~Steinwart and A.~Christmann.
\newblock {\em Support Vector Machines}.
\newblock Springer Science \& Business Media, 2008.

\bibitem{rahimi2007random}
A.~Rahimi and B.~Recht.
\newblock Random features for large-scale kernel machines.
\newblock {\em Advances in neural information processing systems}, 20, 2007.

\bibitem{havlivcek2019supervised}
V.~Havl{\'\i}{\v{c}}ek et~al.
\newblock Supervised learning with quantum-enhanced feature spaces.
\newblock {\em Nature}, 567(7747):209--212, 2019.

\bibitem{kobayashi2003quantum}
H.~Kobayashi, K.~Matsumoto, and T.~Yamakami.
\newblock Quantum merlin-arthur proof systems: Are multiple merlins more
  helpful to arthur?
\newblock In {\em International Symposium on Algorithms and Computation}, pages
  189--198. Springer, 2003.

\bibitem{2020SciPy-NMeth}
P.~Virtanen et~al.
\newblock {{SciPy} 1.0: Fundamental Algorithms for Scientific Computing in
  Python}.
\newblock {\em Nature Methods}, 17:261--272, 2020.

\bibitem{preskill2018quantum}
J.~Preskill.
\newblock Quantum computing in the nisq era and beyond.
\newblock {\em Quantum}, 2:79, 2018.

\bibitem{schuld2021supervised}
M.~Schuld.
\newblock Supervised quantum machine learning models are kernel methods.
\newblock {\em arXiv preprint arXiv:2101.11020}, 2021.

\end{thebibliography}
\bibliographystyle{unsrt}

\end{document}